\documentclass[useAMS,usenatbib]{mn2e}
\usepackage[latin2]{inputenc} % Accept different input encodings
\usepackage[pdftex]{graphicx} %  Enhanced support for graphics
\usepackage{hyperref} %  used to handle cross-referencing commands 
\usepackage{subcaption} % Support for sub-captions ..
\usepackage{amsmath} % improving printed output of documents containing mathematical formulas
\usepackage{amssymb} % internally loads amsfonts
\usepackage{float} % Improves the interface for defining floating objects such as figures and tables
\usepackage{pdflscape}
\usepackage{lipsum}
\usepackage{color,soul} % provides colour highlighting (and management)

\graphicspath{{./figs/}}
\newcommand{\HI}{H\,{\sc i} }
\newcommand{\Htwo}{H$_{2}$ }
\newcommand{\HInospace}{H\,{\sc i}}

\newcommand{\Mstar}{M$_{\star}$ }
\newcommand{\squiggle}{$\mathtt{\sim}$}
\newcommand{\mustar}{$\mu_{\star}$ }

\title[Structure, Star Formation and Gas Content]
{The Effect of Structure and Star Formation on the Gas Content of Nearby Galaxies}
\author[Toby Brown et al.]
{Toby Brown$^1$\thanks{E-mail:
thbrown@swin.edu.au},
Barbara Catinella$^1$,
Luca Cortese$^1$, 
Virginia Kilborn$^1$,
\newauthor
Martha P. Haynes$^2$ and
Riccardo Giovanelli$^2$
\\
$^1$Centre for Astrophysics and Supercomputing, Swinburne University of Technology, Hawthorn, VIC 3122, Australia\\
$^2$Centre for Radiophysics and Space Research, Space Sciences Building, Cornell University, Ithaca, NY 14853, USA}

\begin{document}
\date{Accepted 2015 June 06. Received 2015 June 04; in original form 2015 May 10}

\pagerange{\pageref{firstpage}--\pageref{lastpage}} \pubyear{2002}

\maketitle

\label{firstpage}

\begin{abstract}
We revisit the main \HInospace-to-stellar mass ratio (gas fraction) scaling relations, taking advantage of the HI spectral stacking technique to understand the dependence of gas content on the structural and star formation properties of nearby galaxies. This work uses a volume-limited, multi-wavelength sample of \squiggle25,000 galaxies, selected according to stellar mass ($10^{9} \; \text{M}_\odot \leq \; \text{M}_{\star} \leq 10^{11.5} \; \text{M}_\odot $) and redshift ($0.02 \leq {\it z} \leq 0.05 $) from the Sloan Digital Sky Survey, and with \HI data from the Arecibo Legacy Fast ALFA survey. We bin according to multiple parameters of galaxies spanning the full gas-poor to -rich regime in order to disentangle the dominance of different components and processes in influencing gas content. For the first time, we show that the scaling relations of gas fraction with stellar mass and stellar surface density are primarily driven by a combination of the underlying galaxy bimodality in specific star formation rate and the integrated Kennicutt-Schmidt law. Finally, we produce tentative evidence that the timescales of \HI depletion are dependent upon galaxy mass and structure, at fixed specific star formation rate.

\end{abstract}


\section{Introduction}
\label{sec:Introduction}
Interstellar neutral atomic hydrogen (\HInospace) is the principle gaseous component of galaxies and supplies the raw fuel for star formation. As a result, understanding the properties of \HI is critical if we are to form a complete picture of galaxy evolution. There are strong correlations and complex relationships between galaxy gas content and the properties of colour, shape, size and star formation \citep[see reviews by][and references therein]{Kennicutt1998,Boselli2006}. While it is known that these parameters are regulated by the processes that govern galaxy evolution and the timescales over which they occur, the extent of the interplay between global galaxy properties and the cold gas component remains unclear. 

Until recently, large area comprehensive surveys of extragalactic \HI were non-existent. This meant that studies of cold gas lagged far behind those of stellar populations in both statistics and detail. However, over the past decade, 21-cm line observations have dramatically increased the quantity and quality of data available, characterising the neutral hydrogen content for thousands of galaxies. The \HI Parkes All-Sky Survey \citep[HIPASS;][]{Meyer2004} contains \squiggle5,000 \HInospace-selected galaxies detected in emission out to \squiggle180 Mpc and was the first blind \HI survey to provide such a sample. Its successor, the Arecibo Legacy Fast ALFA \citep[ALFALFA;][]{Giovanelli2005a} survey is measuring \HI emission over $7000\;\text{deg}^2$ of sky out to a distance of \squiggle250 Mpc, reaching nearly eight times better sensitivity, four times better angular resolution and three times the spectral resolution of HIPASS. The latest data release of ALFALFA covers forty per cent of the total survey area and contains 15,855 individually detected sources \citep[$\alpha$.40 H I Source Catalog;][]{Haynes2011}. In addition to blind surveys, deep observations have measured global and resolved \HI properties for a smaller yet growing number of galaxies \citep[e.g.][]{vanderHulst2001,Walter2008}. The largest targeted census of \HI for a representative sample to date is the GALEX Arecibo SDSS Survey \citep[GASS;][]{Catinella2010}, providing \HI quantities for a stellar mass selected sample of \squiggle800 nearby galaxies with stellar masses greater than $10^{10} \; \text{M}_{\odot}$.

At other wavelengths, large scale spectroscopic and photometric surveys such as the Galaxy Evolution Explorer \citep[GALEX;][]{Martin2005,Morrissey2007} surveys in the UV, the Sloan Digital Sky Survey \citep[SDSS;][]{York2000} in the optical, the Two-Micron All Sky Survey \citep[2MASS;][]{Skrutskie2006} in the near-infrared and the Wide-field Infrared Survey Explorer \citep[WISE;][]{Wright2010} cover a significant fraction of the sky and provide a plethora of multi-wavelength data for $10^5-10^6$ objects. Combining such survey efforts across the entire breadth of the electromagnetic spectrum allows for an increasingly in-depth and statistical approach to the study of galaxy populations and their properties. 

Sensitivity limitations of current radio telescopes mean that direct detection of \HI emission for large numbers of galaxies is limited to the local Universe \citep{Catinella2015}. Deep targeted observations probe lower gas content effectively but are exceptionally time expensive and as such are unable to provide samples capable of statistically matching the scope and scale of the UV, optical and IR surveys. On the other hand, blind \HInospace-surveys cover huge volumes and large numbers of galaxies but are not sensitive enough to yield detections for the lower end of the \HI mass function beyond the very closest systems.

However, we may exploit \HInospace-blind surveys beyond their nominal sensitivity limits. Taking advantage of the spectral stacking technique, we compute average \HInospace-to-stellar mass ratios (gas fractions) from high quality 21-cm line spectra of galaxies that are not necessarily individually detected in \HInospace, yet selected using optical coordinates and redshifts. Ultimately, until next generation facilities such as the Square Kilometre Array (SKA) survey vast areas of sky with unprecedented speed and sensitivity, the stacking of \HI spectra for an ensemble of detected and undetected galaxies is the best (and only) way to accurately quantify the gas fraction for such large, representative samples of galaxies. The method has been previously employed with great success to probe lower gas content in both emission \citep{Fabello2011a,Fabello2012,Gereb2013} and absorption \citep{Gereb2014}, and to estimate the \HI cosmic density at intermediate redshift \citep[e.g.][]{Lah2007,Rhee2013}.

There have been many studies into the relationships of gas content with star formation and galaxy structure for thousands of systems. \citet{Fabello2011a}  applied \HI stacking to a large multiwavelength dataset, quantifying the main scaling relations of gas fraction with stellar mass, stellar surface density and colour for \squiggle5,000 massive galaxies. Works such as \citet{Kannappan2004}, \citet{Cortese2011}, \citet{Oh2011} and \citet{Catinella2013} have looked at the same relations using targeted, deep observations. All these investigations showed a strong negative trend of gas content with stellar mass, stellar surface density and colour, identifying stellar surface density and colour, respective morphological and star formation indicators, as the two properties most tightly correlated with gas fraction.

Such studies highlight the role of internal structure and star formation in the regulation of atomic gas content. Understanding the extent and causality of this relationship is of vital importance and has been the focus of much recent work \citep[e.g.][]{Fabello2011a,Cortese2011,Catinella2013}. However, the processes are not as yet fully understood and studies have frequently arrived at differing conclusions, either downplaying the importance of a bulge or bar in affecting gas content or, contrastingly, suggesting such structures may have an influence upon gas consumption \citep[e.g.][]{Saintonge2012,Leroy2013,Huang2014a}.

In this paper we use stacking analysis with unprecedented statistics to build comprehensive scaling relations with gas content, investigating the {\it separate} influences of mass, structure and star formation on cold gas for the entire gas-poor to -rich regime.

The paper is organised as follows. Section \ref{sec:Sample} describes the sample selection criteria and multi-wavelength data used in this work. A brief overview of the \HI spectral stacking technique and the errors involved is given in Section \ref{sec:stacking}. The results, scaling relations and analysis are presented in Section \ref{sec:ScalingRelations} and, finally, we compare our work with previous studies and discuss implications in the context of galaxy evolution in Section \ref{sec:Discussion}.

Throughout this paper the distance dependent quantities are computed assuming a $\Lambda$CDM cosmology with $\Omega = 0.3$, $\Lambda = 0.7$, and a Hubble constant $\text{H}_{0} = 70\; \text{km s}^{-1} \;\text{Mpc}^{-1}$.

\section[]{The Sample}
\label{sec:Sample}
\subsection{Selection Criteria}
\label{sec:SampleSelections}
We define a volume-limited, stellar mass selected {\it parent sample} of  30,695 galaxies from the intersection of the Sloan Digital Sky Survey Data Release 7 \citep[SDSS DR7;][]{Abazajian2009} footprint and sky region for which ALFALFA data have been processed to date.

The stellar mass, M$_{\star}$, and volume cuts applied are as follows:

\begin{center}
$10^{9} \; \text{M}_\odot \leq \; \text{M}_{\star} \leq 10^{11.5} \; \text{M}_\odot $
\\$112.5\textdegree \leq \alpha \leq 247.5\textdegree $
\\$0\textdegree \leq \delta \leq 18\textdegree $ \& $24\textdegree \leq \delta \leq 30\textdegree $
\\$0.02 \leq {\it z} \leq 0.05 $
\end{center}
Stellar mass estimates, derived via photometric fitting, are taken from the MPA-JHU SDSS DR7
\footnote{{\it http://www.mpa-garching.mpg.de/SDSS/DR7/}

In this work we use the improved stellar masses from: {\it http://home.strw.leidenuniv.nl/\squiggle jarle/SDSS/}} 
catalogue. The sample straddles the `transition mass' in galaxy populations of $3 \times 10^{10} \; \text{M}_{\odot}$ identified by \citet{Kauffmann2003}.

The significant contribution of peculiar velocities to galaxy redshift measurements is avoided by setting a lower redshift limit of {\it z} $\geq$ 0.02 while the frequency range over which the San Juan airport radar affects Arecibo's observation of redshifted \HI emission is removed by the ceiling of {\it z} $\leq$ 0.05.

The sample also lies in the sky footprint of the GALEX surveys. Therefore, for this work, the sample is further reduced in size because we include only those galaxies detected in near-ultraviolet (NUV, see Section \ref{sec:SDSSGALEX} for details). Finally, we discard an additional one per cent of the sample due to radio frequency interference (RFI) contamination of ALFALFA \HI spectra. A full description of the SDSS optical, GALEX NUV and ALFALFA \HI data is given in sections {\ref{sec:SDSSGALEX}} and {\ref{sec:ALFALFA}}.

\begin{figure*}
\centering
  \captionsetup[subfigure]{labelformat=empty}
   \begin{subfigure}[h]{0.495\linewidth} \centering
     \includegraphics[width=90mm,scale=1.5]{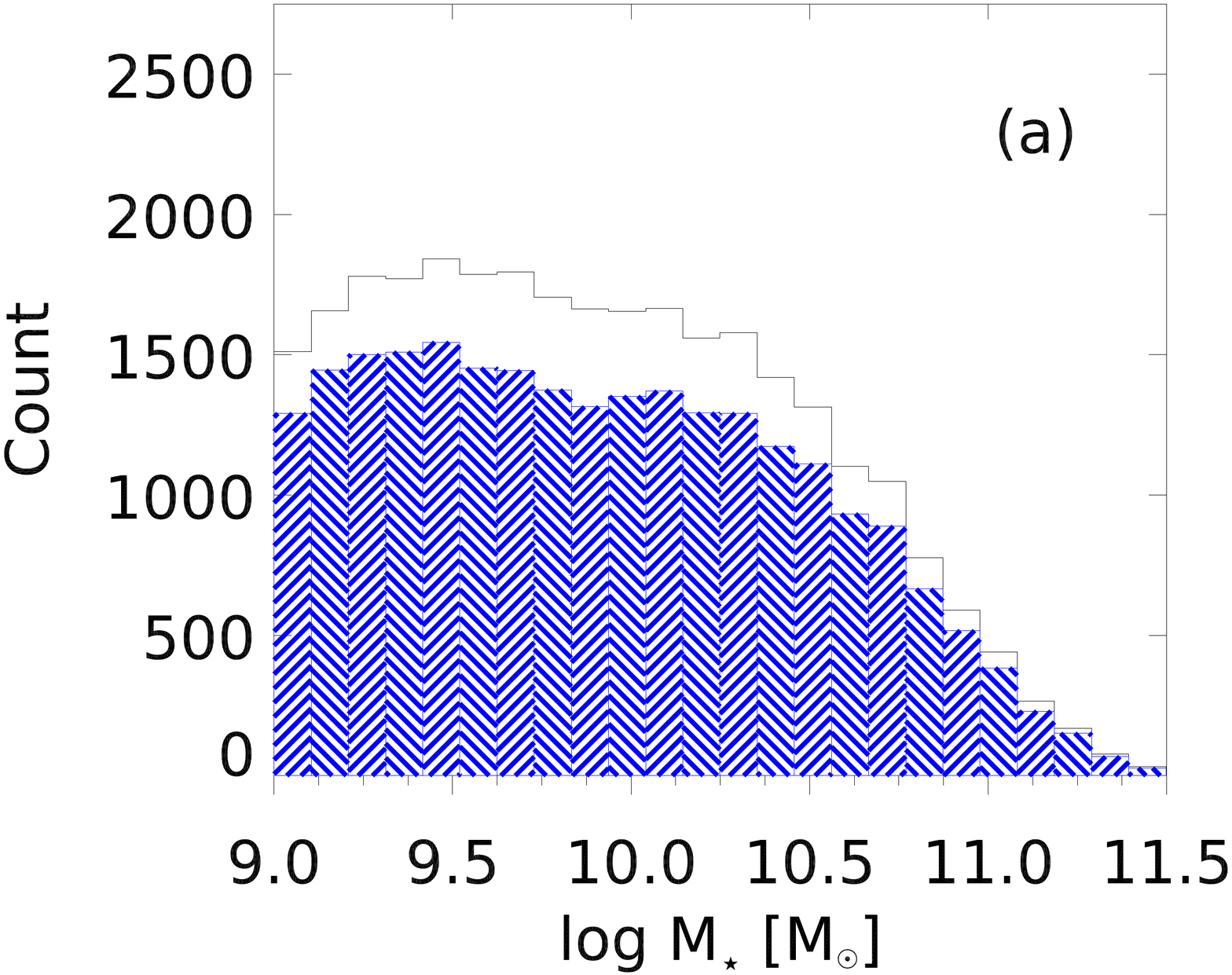}\hfill%
     \caption{}\label{fig:Mst_hist}
   \end{subfigure}
   \begin{subfigure}[h]{0.495\linewidth} \centering
     \includegraphics[width=90mm,scale=1.5]{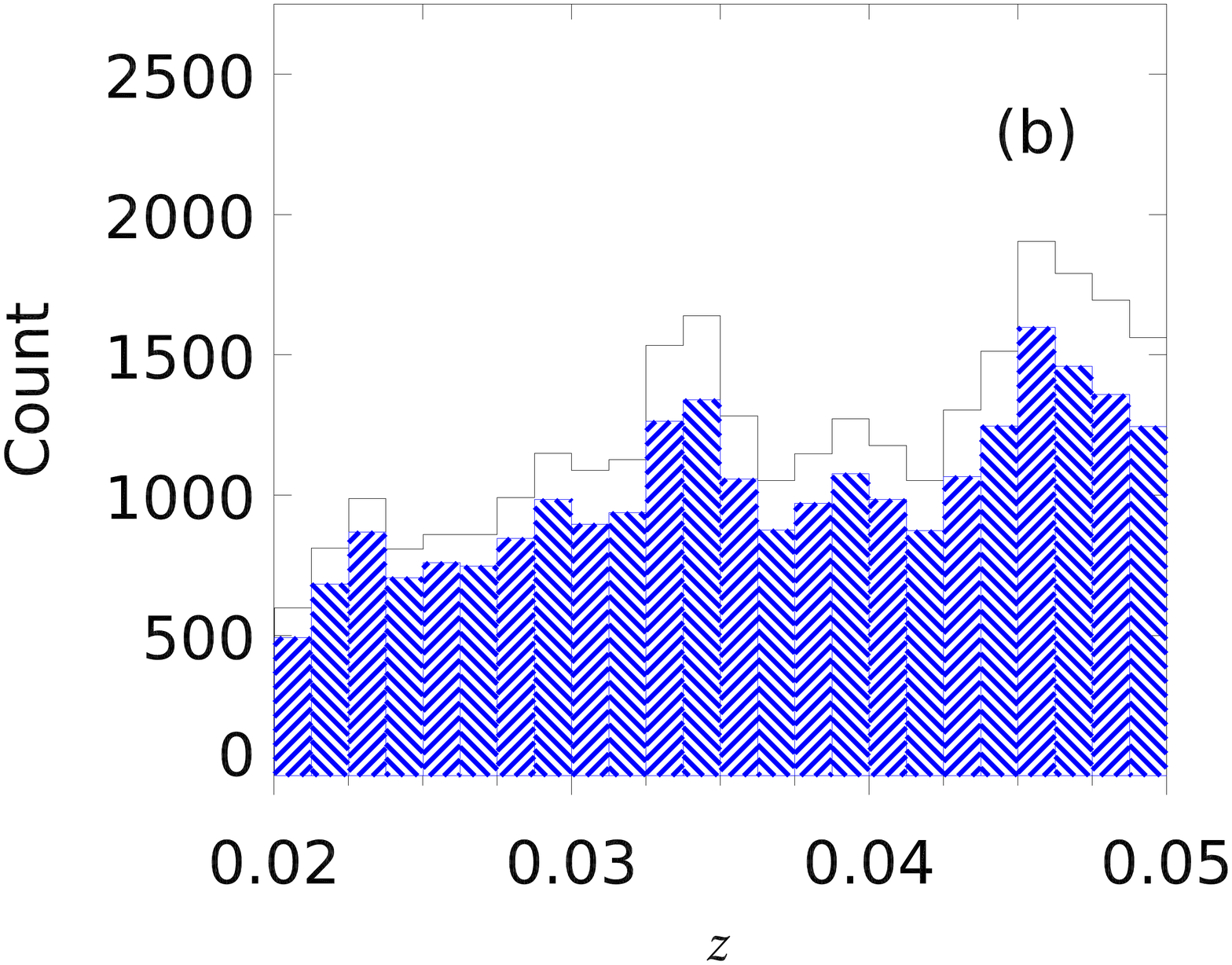}\hfill%
     \caption{}\label{fig:Z_hist}
   \end{subfigure}
   \begin{subfigure}[h]{0.495\linewidth} \centering
     \includegraphics[width=90mm,scale=1.5]{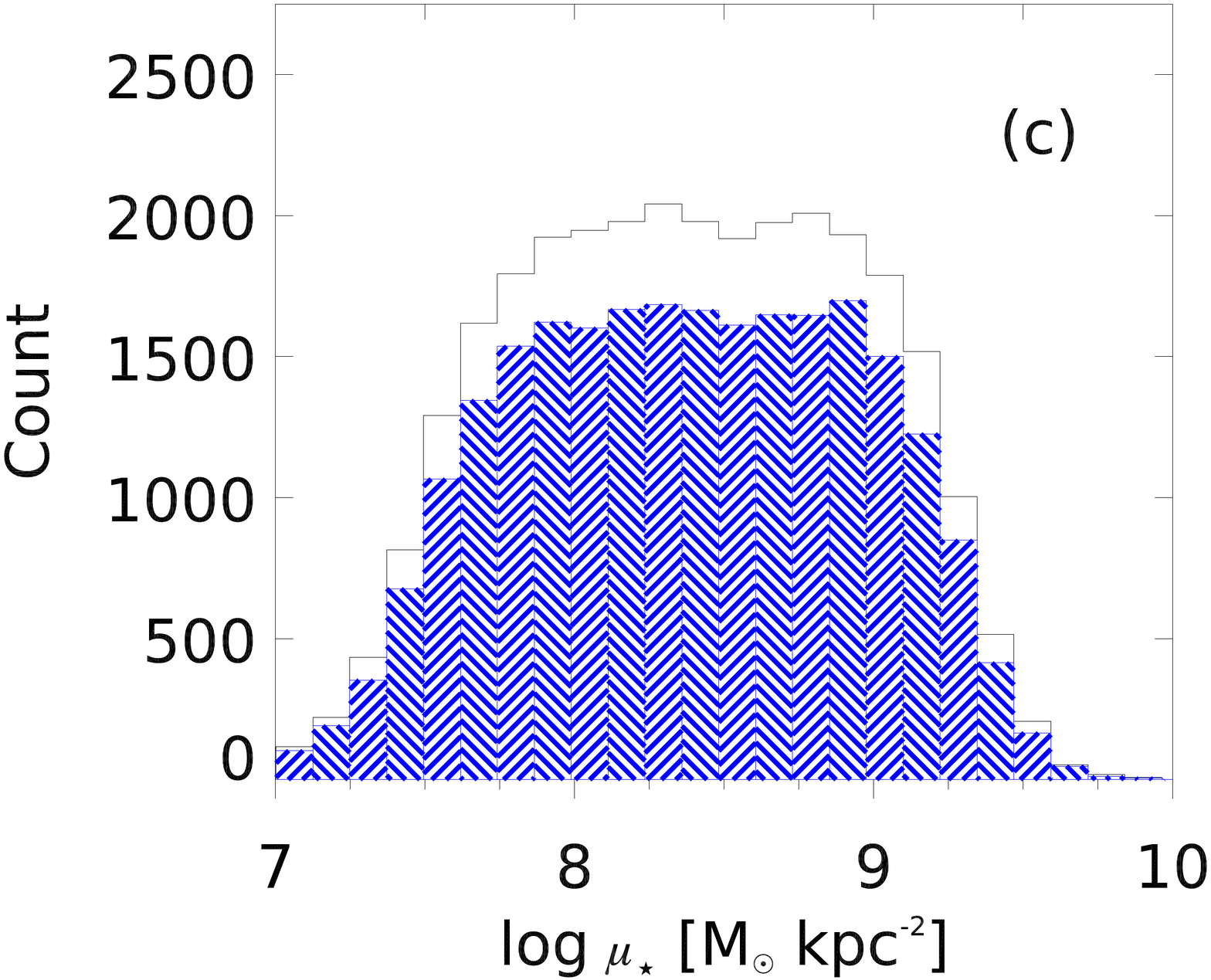}\hfill%
     \caption{}\label{fig:MUst_hist}
   \end{subfigure}
   \begin{subfigure}[h]{0.495\linewidth} \centering
     \includegraphics[width=90mm,scale=1.5]{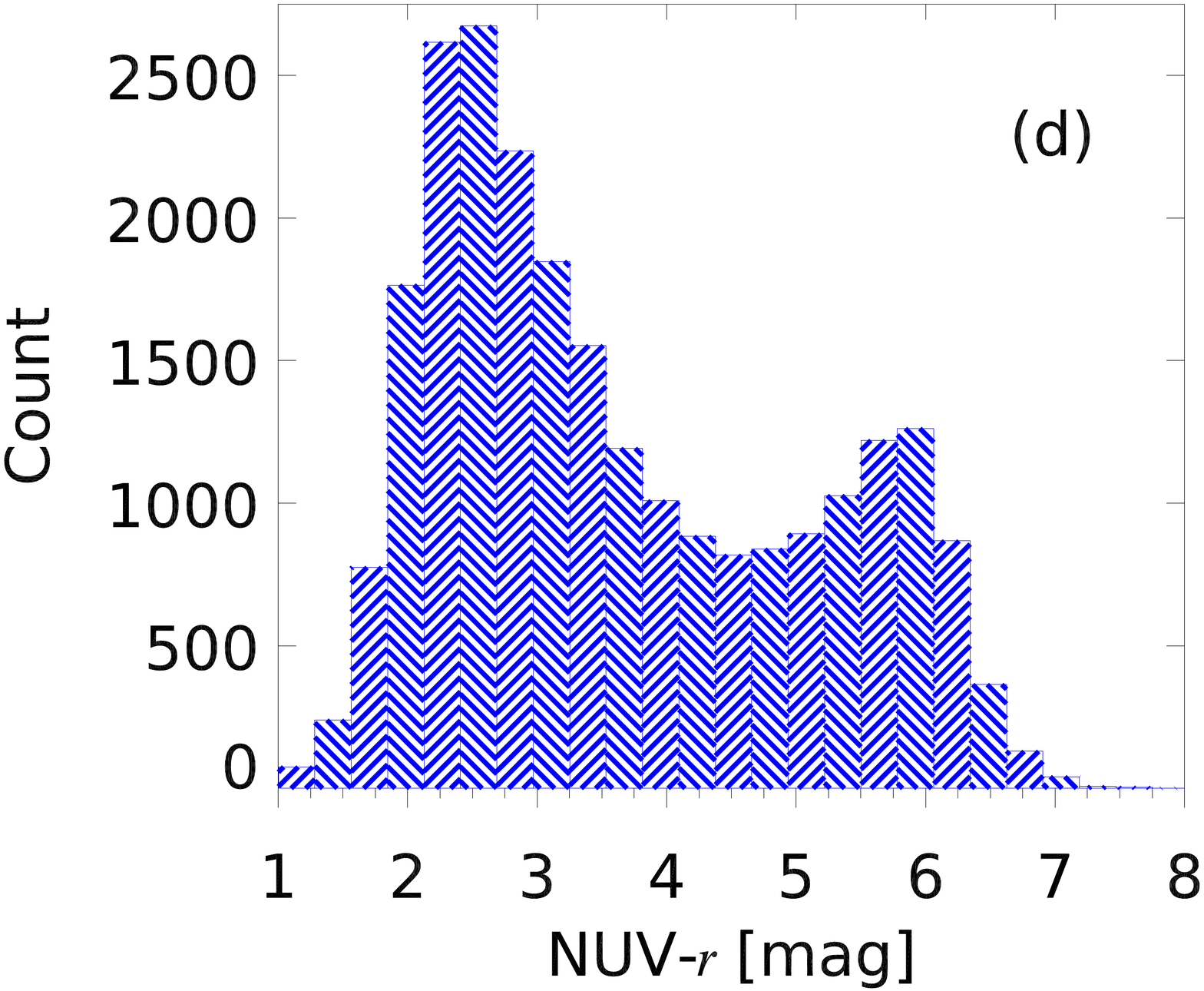}\hfill%
     \caption{}\label{fig:NUVr_hist}
   \end{subfigure}
\caption{Distribution of sample galaxies across stellar mass, redshift, stellar surface density and NUV-{\it r} colour. The black histograms show the parent sample; the blue, dashed histograms represent the final sample of 24,337 galaxies used in this paper.}
\label{fig:sampledist}
\end{figure*}

The total sample of SDSS DR7 galaxies with high quality (i.e. uncontaminated by RFI) ALFALFA spectra and GALEX NUV photometry is 24,337 galaxies.

\subsection{Optical and UV Data}
\label{sec:SDSSGALEX}

Galaxy properties including redshift, size and magnitudes are obtained from the SDSS DR7 database server {\it CasJobs}\footnote{{\it http://skyserver.sdss3.org/casjobs/}} via Structured Query Language queries. Optical photometric data are taken from five broad bands: {\it u, g, r, i, z} \citep[AB system;][]{Fukugita1996}. For magnitudes and colours we take the `model' magnitudes. These are the optimal fit of either a pure de Vaucouleurs or a pure exponential profile to the galaxy flux in each band. We compute distance measurements and rest-frame shift \HI spectra using fibre spectroscopic redshifts.

The UV data are obtained by carefully cross-matching the GALEX  Unique Source Catalogues\footnote{\it http://archive.stsci.edu/prepds/gcat/} \citep[GCAT;][]{Seibert2012} and the Bianchi, Conti, Shiao Catalogue of Unique GALEX Sources\footnote{\it http://archive.stsci.edu/prepds/bcscat/} \citep[BCS;][]{Bianchi2014} to the SDSS database using an impact parameter of ten arcseconds. The two UV catalogues use very similar techniques and as such the difference between their photometries is small ($\leq$ 0.1 mag). Where galaxies are present in both catalogues we selected magnitudes from the GCAT as this has a larger overlap with our sample. GCAT does not include the GALEX GR7 data release and so, when necessary, we also draw values from the BCS catalogue. We use magnitudes obtained from the Medium Imaging Survey \citep[MIS; see ][]{Martin2005} images for the ninety-six per cent of the sample where it is available, and from the All-Sky Imaging Survey (AIS) for the remaining four per cent.

All optical photometric data are corrected for Galactic extinction following \citet*{Schlegel1998} and using the respective band extinction obtained from the SDSS DR7 {\it PhotObj} catalogue. Reddening correction for NUV photometry is according to \citet{Wyder2007}, who adopt $\text{A}(\lambda)/\text{E(B-V)} = 2.751$ for SDSS {\it r}-band and $\text{A}(\lambda)\text{/E(B-V)} = 8.2$ for GALEX NUV. Thus we convert to NUV extinction, $\text{A}_{\text{NUV}}$, using $\text{A}_{\text{NUV}} -  \text{A}_{\it r} = 1.9807\text{A}_{\it r}$, where $\text{A}_{\it r}$ is the {\it r}-band extinction.

Figure \ref{fig:sampledist} shows the distributions of stellar mass, redshift, NUV-{\it r} colour and stellar mass surface density for the galaxies in our sample. We derive the stellar surface density, \mustar, as:
\begin{equation}
\mu_{\star} = \dfrac{\text{M}_{\star}}{2 \pi \text{R}_{50,{\it z}}^2}
\label{eq:mu_star}
\end{equation}
where $\text{R}_{50,{\it z}}$ is the Petrosian radius containing fifty per cent of the flux in {\it z} band, expressed in kpc.

The distributions of the main optical- and UV-derived parameters for the parent sample (solid black line) and of the galaxies used in this work (hatched blue histograms) are given in Figure \ref{fig:sampledist}. The ratio between sample galaxies with GALEX data used in this work and their parent sample is similarly uniform in distribution across the four properties. This illustrates that, when removing galaxies for which there are no valid NUV magnitudes, we do not inadvertently introduce bias by over or under sampling a region of the parameter space.

Figure \ref{fig:NUVr_hist} is the NUV-{\it r} colour histogram for the sample and exhibits the well known blue cloud and red sequence bimodality \citep{Wyder2007,Schiminovich2007}. The NUV-{\it r} colour is chosen as it probes young and old stellar populations on either side of the 4000\AA \,break and thus is sensitive to the star formation properties of a galaxy.

\begin{figure}
\centering
\includegraphics[width=90mm,scale=1.5]{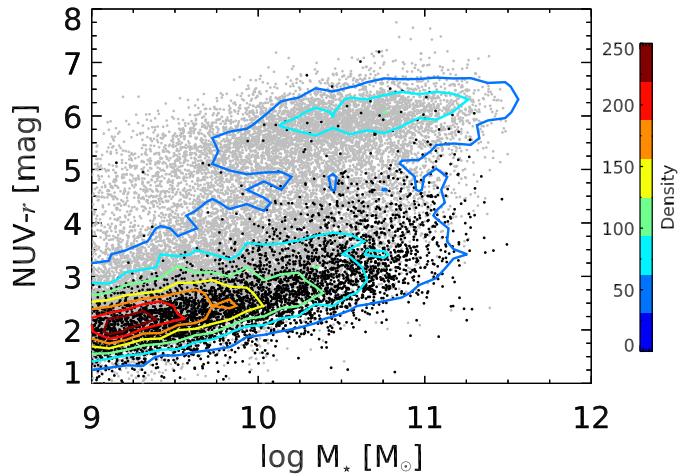} % Include the image
\caption{NUV-{\it r} colour versus stellar mass diagram for galaxies in our sample. Contours represent density levels from a minimum of 20 galaxies per bin up to 250 per bin. The points show positions in the plane for the full sample of 24,337 galaxies. Black points are the 4,610 galaxies detected in \HI by ALFALFA, grey points are undetected galaxies in the same \HI observations.}
\label{fig:UVColourmass}
\end{figure}

\subsection{\HI 21-cm Line Data}
\label{sec:ALFALFA}
The 21-cm \HI spectra are extracted from the full volume of ALFALFA data cubes using positional coordinates, $\alpha$, $\delta$ and redshift, {\it z}, drawn from the SDSS DR7 database. The technique of optical selection from data cubes within the shared footprint of SDSS and ALFALFA provides spectra with excellent baseline characteristics for both detections and non-detections in \HInospace, each with optical counterpart information. For examples of \HI detected and non-detected ALFALFA spectra as well as a full description of the extraction procedure see Section 3 of \citet{Fabello2011a}. 

We extract spectra with a raw spectral resolution of \squiggle5.5 km s$^{-1}$ and spanning \squiggle5,500 km s$^{-1}$ in velocity range (25 MHz in frequency) from a 4 $\times$ 4 arcminute aperture centred on the position of the target galaxy. The quality of each radio spectrum has been conservatively assessed by an automated routine which flagged spectra with spikes in the signal (more than ten times the root-mean-square, rms, value) within a central $1000$ km s$^{-1}$ interval centred on the redshift of the galaxy. We conducted a visual inspection of all spectra that failed this test and discarded those with central RFI.

Figure \ref{fig:UVColourmass} shows the NUV-{\it r} colour-mass diagram for our sample with individual ALFALFA detections and non-detections shown by black and grey points respectively. The superimposed galaxy number density contours show the classic bimodality, separating the blue, star forming from the red, quiescent galaxies. As can be seen, blind surveys such as ALFALFA preferentially detect blue cloud objects (NUV-{\it r} $\leq$ 3.5) and though red sequence detections do occur, they are rare, a result already noted \citep[e.g.][]{Huang2012}.

Galaxies in our sample have Petrosian radii containing ninety per cent of the {\it r}-band flux ({$\text{R}_{90,{\it r}}$}) of less than one arcminute, hence their \HI emission is always unresolved (Arecibo beam FWHM = \squiggle3.5 arcminutes).

The ALFALFA \HI detection (S/N $\geq 6.5$) rate for our sample is nineteen per cent (4,610 galaxies). Reliable individual \HI flux densities for detected galaxies not already included in the $\alpha$.40 ALFALFA data release were provided by M.P. Haynes.

\HI masses are computed via the standard formula:
\begin{equation}
\dfrac{\text{M}_{\text{HI}}}{\text{M}_\odot} = 
\left(\dfrac{2.356 \times 10^5}{1 + {\it z}}\right) 
\left(\dfrac{\text{D}_{{\it L(\it z)}}}{\text{Mpc}}\right) ^2 
\left(\dfrac{\text{S}_{21}}{\text{Jy km s}^{-1}}\right)  % Roberts (1963)
\label{eq:HImassEqn}
\end{equation}
where {\it z} is the redshift, $\text{D}_{{\it L(\it z)}}$ is the luminosity distance and $\text{S}_{21}$ is the integrated 21-cm line flux density. Gas fraction is simply taken as M$_{\text{HI}}$/M$_{\star}$.

\section{\HI Spectral Stacking}
\label{sec:stacking}
With over eighty per cent \HI non-detections, the sample used in this work is ideally designed to exploit the capabilities of the stacking technique, allowing us to determine how \HI content varies with galaxy properties. We use an adapted version of the software developed by \citet{Fabello2011a}. A more comprehensive description of the stacking technique can be found in Section 3.2 of that work.

Stacking makes use of the fact that contained within the three dimensional volume of 21-cm data cubes is the \HI emission from galaxies regardless of whether they are formal detections. Using the optical position and redshift, we extract 21-cm line spectra from the cubes at the locations of our sample galaxies.

Each spectrum is shifted to its rest-frame velocity, so that the galaxy is centred at zero velocity, using the SDSS spectroscopic redshift for the corresponding optical counterpart. The mean values of redshift and stellar mass within each bin may not be representative of the individual galaxies, therefore we weight each spectrum by its redshift and stellar mass before stacking in order to transform the radio signal into a `gas fraction spectrum' \citep[see eqn. 5 in][]{Fabello2011a}. \citet{Fabello2011a} find this method is consistent with - and when the bins are large preferable to - co-adding the raw spectra and using the mean redshift to convert the stacked signal into an average \HI mass and subsequent gas fraction. The rest-frame spectra for N galaxies are then weighted by their rms noise, and co-added, or stacked, yielding a total signal for the sample while reducing the {\it rms} as $1/\sqrt{\text{N}}$. Once the number of spectra stacked reaches the threshold N \squiggle 300 non-Gaussian noise begins to dominate, therefore the reduction in rms continues but at a decreased rate \citep{Fabello2011a}.

In some instances (e.g. where statistics in the bins is low or galaxies have red NUV-{\it r} colours) the stacked signal yields a non-detection. When this occurs upper limits are computed assuming a $5\sigma$ signal with a velocity width of $\text{W} = 200 \; \text{km s}^{-1}$ for bins with $\langle \text{M}_{\star}\rangle  \leq 10^{10}\; \text{M}_{\odot}$ and $\text{W} = 300 \; \text{km s}^{-1}$ where $\langle \text{M}_{\star}\rangle  > 10^{10}\; \text{M}_{\odot}$, smoothed to a spectral resolution of $(\text{W}/2) \;\text{km s}^{-1}$. This approach is outlined in \citet{Giovanelli2005a} and is identical to GASS over the same stellar mass range ($\text{M}_{\star} > 10^{10}\; \text{M}_{\odot}$).

\subsection{Errors}
\label{sec:Errors}
Errors on the average gas fractions obtained with stacking are computed using the {\it Delete-a-Group Jacknife} (DAGJK) method \citep{Kott2001}, a statistical estimate of the standard deviation (or {\it rms} error) that incorporates both variance of the estimator and its bias. Based upon the standard {\it Jacknife} procedure formulated by \citet{Tukey1958}, DAGJK iteratively estimates a given population parameter (i.e. \HI gas fraction) while discarding, in turn, a separate random subset of galaxies from the sample. The weighted difference between the mean statistic measurement and each jackknifed statistic is an estimate of that group's influence on the mean value and an indicator of the variance within the dataset.

The DAGJK standard deviation estimator is:

\begin{equation}
\sigma(t)= \sqrt{var(t)} = \sqrt{\dfrac{(\text{R}-1)}{\text{R}}\sum_{{\it r}=0}^{\text{R}} (t_{({\it r})}-\overline{t})^2}  % Kott (2001)
\label{eq:DAGJKstddev}
\end{equation}
where R is the number of replicated estimates, $t_{(r)}$ is the estimation of the population parameter {\it without} the {\it r}th subset and $\overline{t}$ is the mean of the replicated estimates.

When computing the DAGJK error we discard twenty per cent of the sample without replacement on each iteration, meaning that the number of estimates, R, is five.

It should be noted that the errors shown are not indicative of the standard deviation of the underlying distribution of individual gas fractions, this is unknown. Instead the errors show the effect of strong outliers, if present, on the final stacking result. When employed using random rejection the main advantage of DAGJK over the traditional `leave one out' method is the improvement in computational efficiency while maintaining precision in the (nearly) unbiased confidence interval of the population parameter. The result is entirely consistent with the standard Jackknife technique. Additional advantages of DAGJK will be explored in follow up investigations, where we will estimate errors on the stacked average gas fraction by targeted rejection according to galaxy properties (i.e. NUV-{\it r} colour). DAGJK will be capable of giving an indication as to how particular galaxy populations influence the mean gas content by providing some insight into the variance of the stacking average.

\section{\HI Gas Fraction Scaling relations}
\label{sec:ScalingRelations}
In this section we present the main scaling relations of gas fraction versus stellar mass, stellar surface density and NUV-{\it r} colour for our sample, based upon spectral stacking of ALFALFA data. Compared to previous work by \citet{Fabello2011a}, which was based upon a subset of our sample (\squiggle5000 galaxies), we are able to extend our analysis down one order of magnitude in stellar mass ($\text{M}_{\star} \geq 10^9 \text{M}_\odot$) as well as investigate second order trends in gas fraction, gaining further insights into the physical drivers of these relations.

\begin{figure*}%
\centering
  \captionsetup[subfigure]{labelformat=empty}
   \begin{subfigure}{0.51\linewidth}\centering
     \includegraphics[width=87mm,scale=1.5]{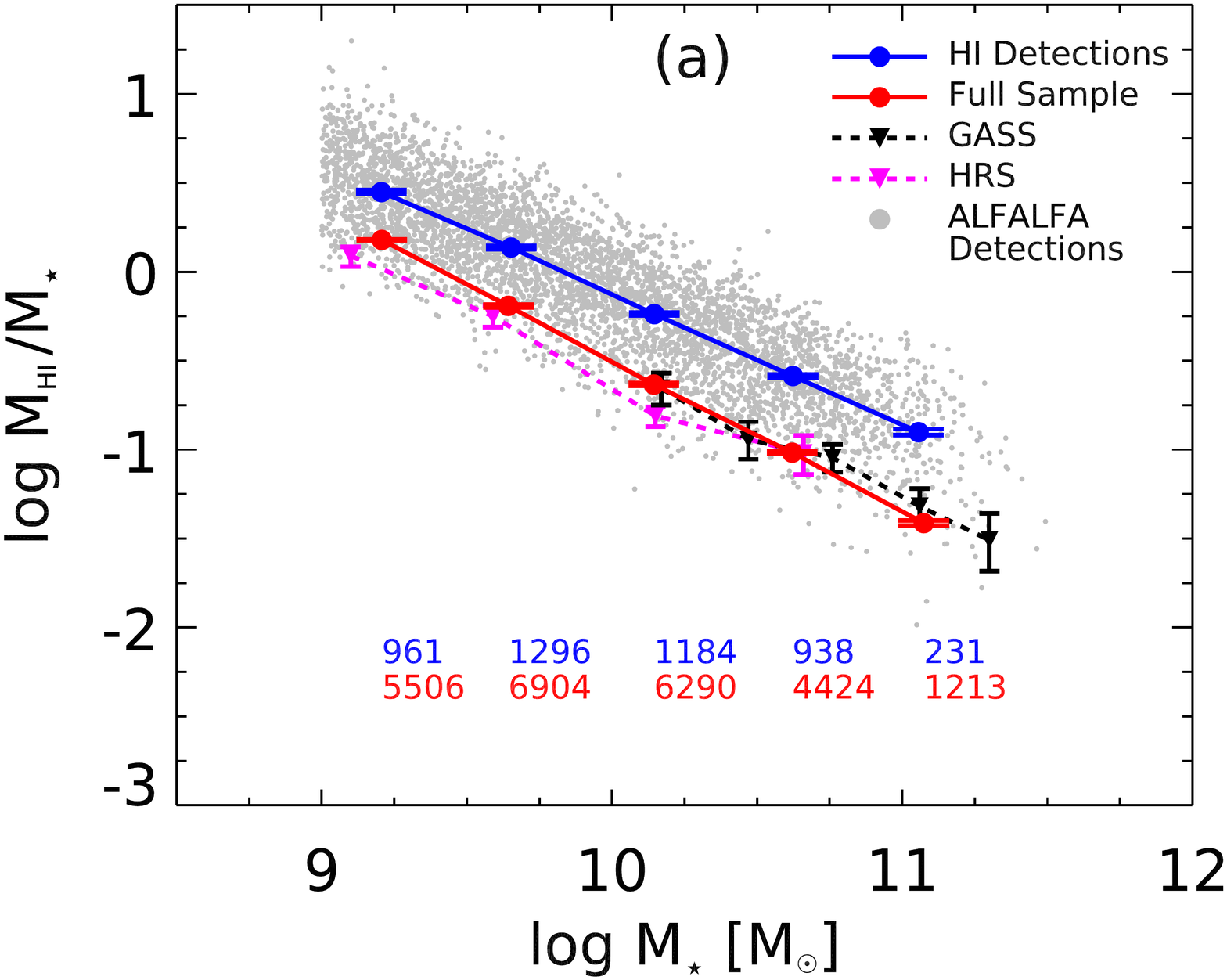}
     \caption{}\label{fig:Mst_det}
   \end{subfigure}
   \begin{subfigure}{0.49\linewidth}\centering
     \includegraphics[width=87mm,scale=1.5]{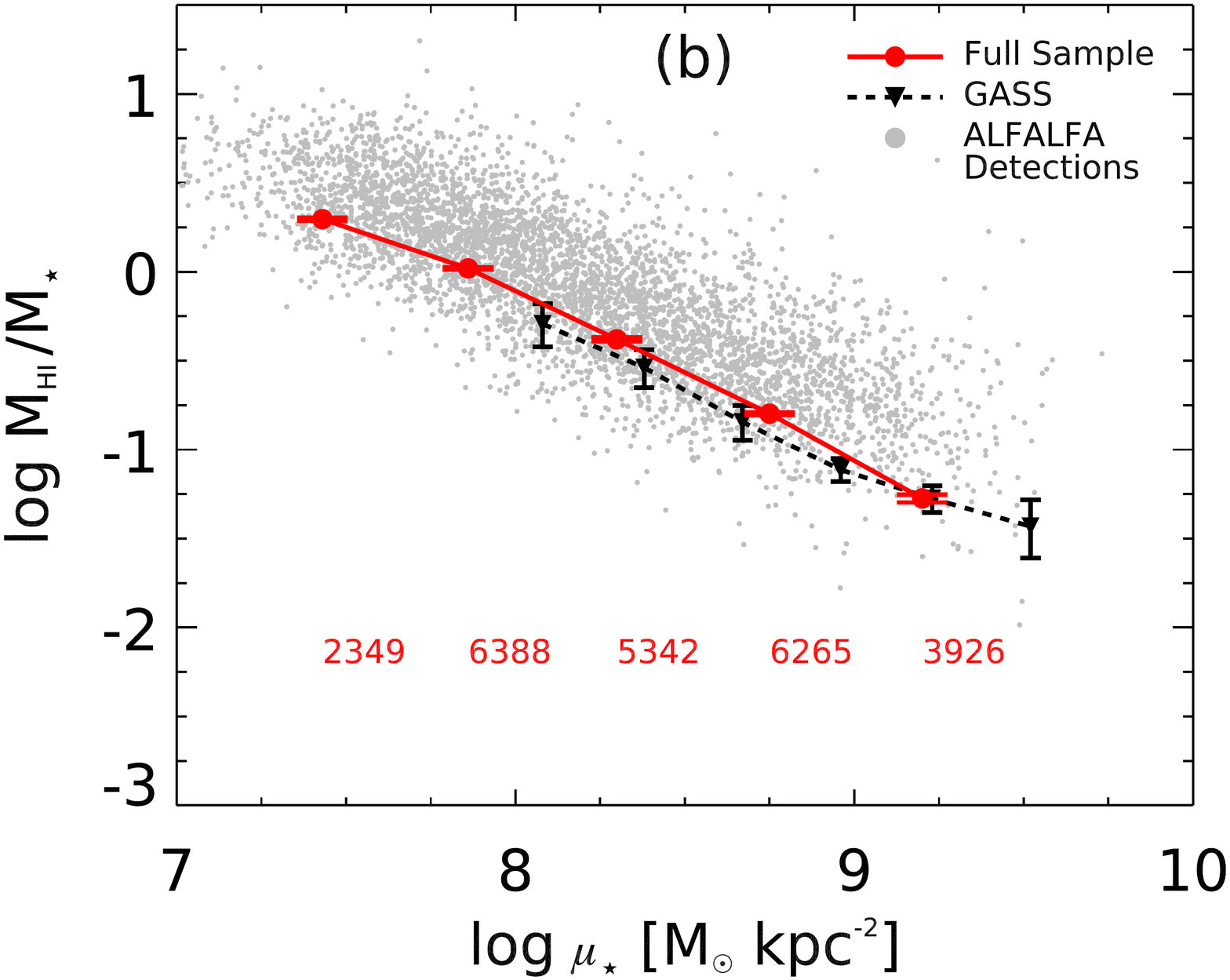}
     \caption{}\label{fig:GF_mu}
   \end{subfigure}
   \begin{subfigure}{0.49\linewidth}\centering
     \includegraphics[width=87mm,scale=1.5]{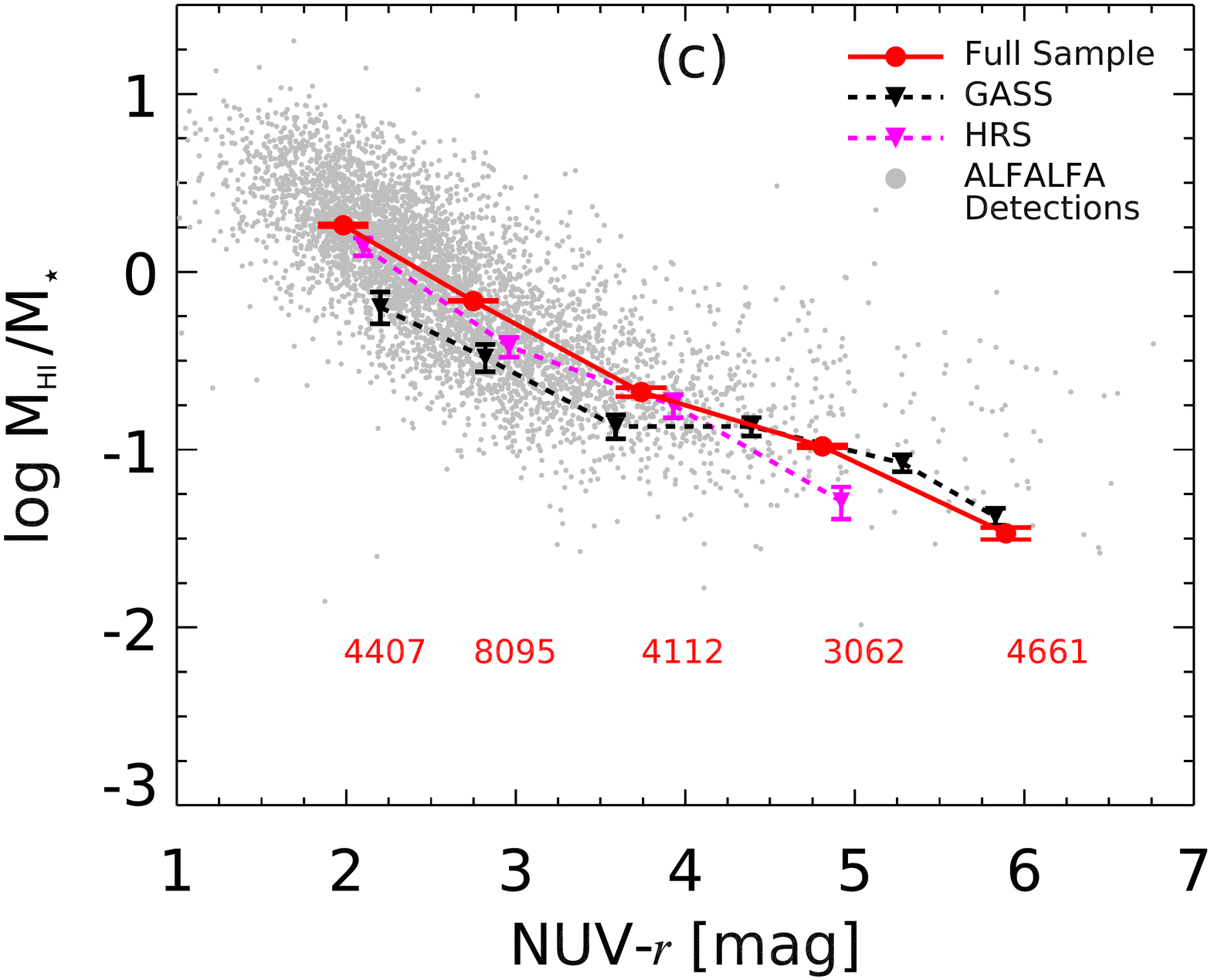}
     \caption{}\label{fig:GF_nuvr}
   \end{subfigure}
\caption{
Average stacked \HI gas fractions are plotted as a function of galaxy stellar mass (M$_{\star}$), stellar surface density ($\mu_{\star}$) and NUV-{\it r} colour for the whole sample. Grey points are individual detections included in the ALFALFA catalogue. The magenta dashed line is average gas fraction of nearby \HI detections from the Herschel Reference Survey (HRS; Boselli et al., 2010). Dashed black line shows the scaling relation from GASS. Numbers represent the total number of co-added galaxies within each bin. The blue line in Figure \ref{fig:Mst_det} is the stacked average of the galaxies that are detected in ALFALFA. The average gas fraction data points for each relation are given in Table \ref{tab:gf_table}.}
\label{fig:main_SR}
\end{figure*}

Figure \ref{fig:main_SR} shows stacked average \HI fraction as a function of galaxy stellar mass, stellar surface density and NUV-{\it r} colour, shown by the solid red lines. Errors on the average \HI fractions are computed using DAGJK, as described in Section \ref{sec:Errors}. Grey points indicate individual \HI detections from ALFALFA. In Figures \ref{fig:Mst_det} and \ref{fig:GF_nuvr} we also plot the average {\it{linear}} gas fractions\footnotemark \citep{Cortese2011}, provided by L. Cortese, from the Herschel Reference Survey \citep[HRS,][]{Boselli2010}, shown by the magenta dashed line. We exclude \HInospace-deficient galaxies - typically found within clusters - because of their significant offset to lower gas content. We do not show the HRS results in Figure \ref{fig:GF_mu} because of a difference in the definitions of stellar surface density between their work and ours. We confirm the trends of decreasing \HI fraction as a function of galaxy stellar mass, stellar surface density and NUV-{\it r} colour, even with the addition of lower stellar mass galaxies. Our results are entirely consistent with the results of \citet{Fabello2011a} as well as the findings of \citet{Catinella2013}, using the log of the linear gas fraction averages from the final GASS data release (dashed black line). The data points for the full sample scaling relations shown in Figure \ref{fig:main_SR} are given in Table \ref{tab:gf_table}.

\footnotetext{The reader should note that the distribution of gas fractions in the local Universe is more likely lognormal than Gaussian, so ideally one would compute $\langle \text{log} \: \text{M}_{\text{HI}}/\text{M}_{\star} \rangle$ \citep{Cortese2011}. However, the stacking method does not operate in log space, instead it returns the linear average of the \HI content, so we must adopt $\text{log} \: \langle \text{M}_{\text{HI}}/\text{M}_{\star} \rangle$ in its place. Care must be taken when comparing our results with gas fractions from deep, detection dominated surveys such as GASS and HRS that the average is taken of the linear data, not of the log-scaled value, as the log of the average is not the average of the log.}

\begin{table}
\caption{Average gas fractions for the full sample scaling relations shown in Figure \ref{fig:main_SR}. The column labelled {\it x} is the galaxy property along the x-axis,  $\langle x \rangle$ denotes the mean values of {\it x} within each bin, $\langle \text{M}_{\text{HI}}/\text{M}_{\star} \rangle$ is the linear gas fraction and N gives the number of galaxies within each bin.}
\label{tab:gf_table}
\centering
\begin{tabular}{lrll}
\hline
{\it x}    & $\langle x \rangle$    & $\langle \text{M}_{\text{HI}}/\text{M}_{\star} \rangle$    &   N\\
\hline
log M$_{\star}$       &   9.21    &  1.511 $\pm$ 0.011  &  5506 \\
                                 &   9.64    &  0.643 $\pm$ 0.011  &  6904 \\
                                 &   10.14  &  0.232 $\pm$ 0.005  &  6290 \\
                                 &   10.62  &  0.096 $\pm$ 0.002  &  4424 \\
                                 &   11.07  &  0.039 $\pm$ 0.001  &  1213 \\
                                 &              &                                  &          \\
log $\mu_{\star}$    &   7.43  &   1.976 $\pm$ 0.041  &  2349 \\
                                 &   7.86  &  1.045 $\pm$ 0.014  &  6388 \\
                                 &   8.30  &  0.417 $\pm$ 0.010  &  5342 \\
                                 &   8.75  &  0.159 $\pm$ 0.002  &  6265 \\
                                 &   9.20  &  0.053 $\pm$ 0.003  &  3926 \\
                                 &              &                                  &          \\
NUV-{\it r}              &   1.98  &   1.827 $\pm$ 0.029  &  4407 \\
                                 &   2.75  &  0.685 $\pm$ 0.005  &  8095 \\
                                 &   3.74  &  0.211 $\pm$ 0.012  &  4112 \\
                                 &   4.81  &  0.104 $\pm$ 0.002  &  3062 \\
                                 &   5.89  &  0.034 $\pm$ 0.003  &  4661 \\
\hline
\end{tabular}
\end{table}

The blue line in Figure \ref{fig:Mst_det} shows the result of stacking only galaxies that are detected by ALFALFA, corroborating previous findings that \HI selected samples, unless corrected, will overestimate the average gas content of galaxies within the volume \citep[see Figure 2 in][]{Huang2012}.

\begin{figure*}
\centering
  \captionsetup[subfigure]{labelformat=empty}
   \begin{subfigure}{0.49\linewidth} \centering
     \includegraphics[width=87mm,scale=1.5]{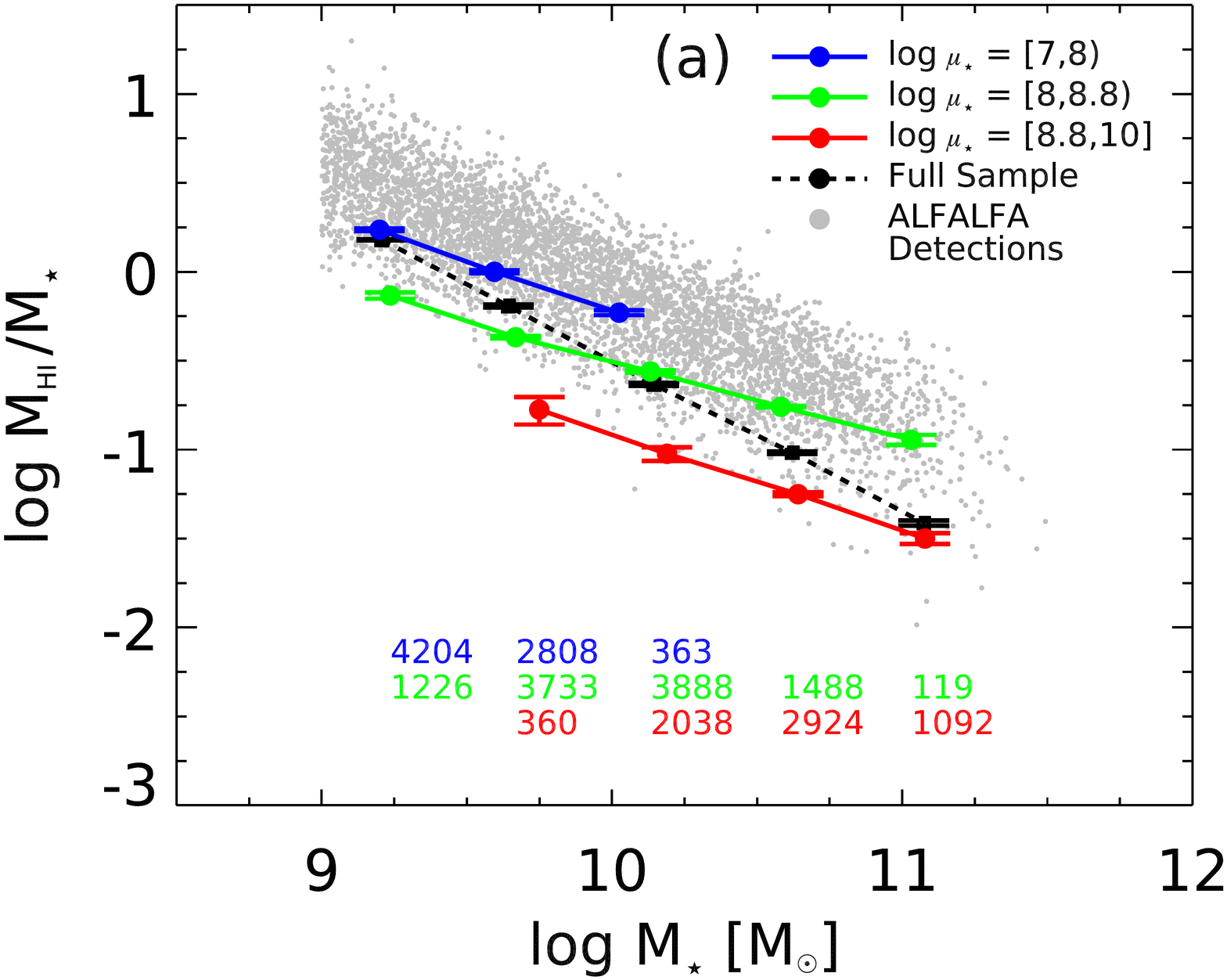}
     \caption{}\label{fig:figAmu}
   \end{subfigure}
   \begin{subfigure}{0.49\linewidth} \centering
     \includegraphics[width=87mm,scale=1.5]{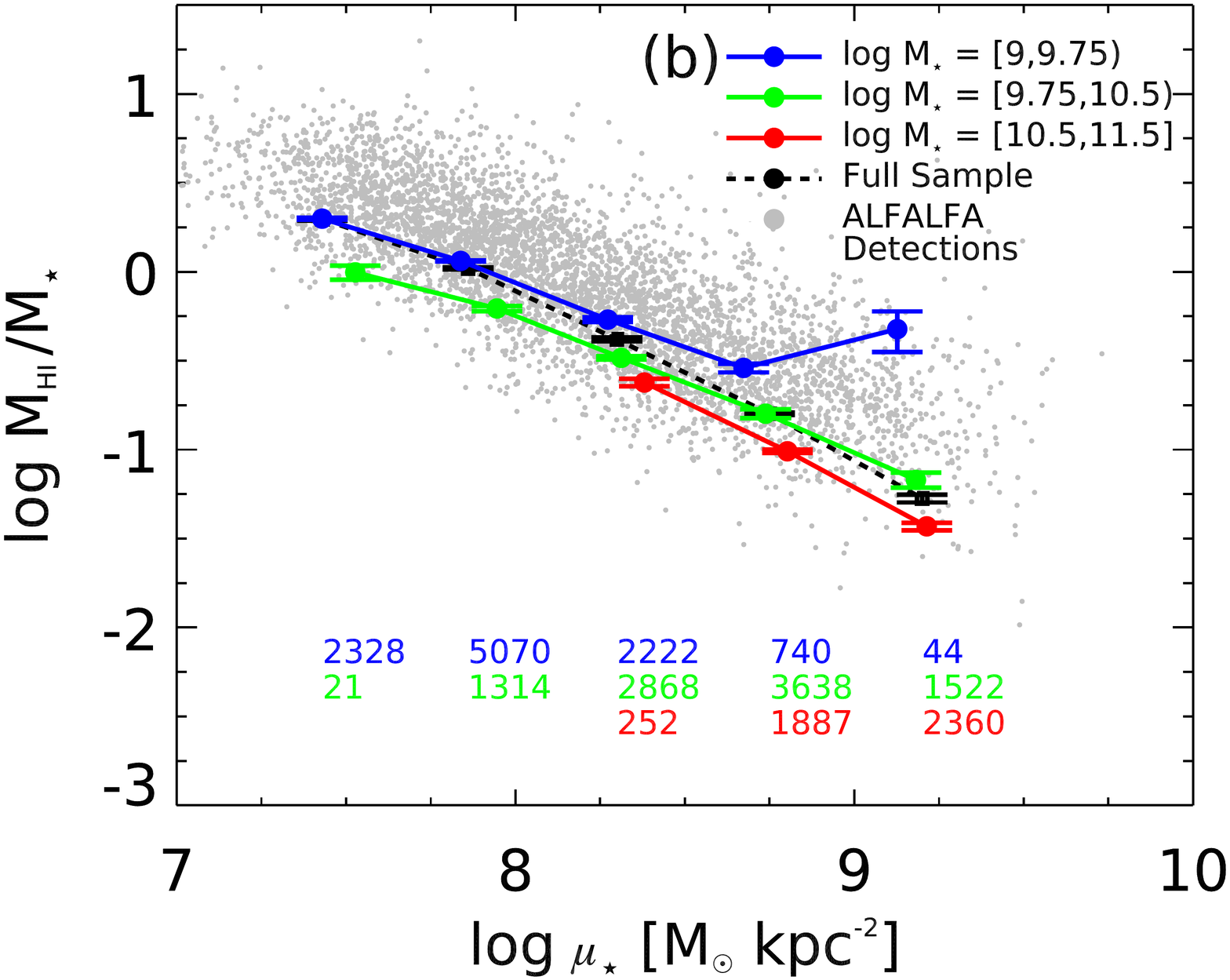}
     \caption{}\label{fig:figBmu}
   \end{subfigure}
\caption{Left: \HI gas fraction as a function of stellar mass, separated into low (blue), intermediate (green) and high (red) stellar surface density bins, as indicated on the top right. Right: \HI gas fraction as a function of stellar surface density, separated into bins of stellar mass. In both panels, grey points indicate ALFALFA detections and the numbers below the relations indicate the number of galaxies co-added in each bin. The dashed black lines show the scaling relations for the whole sample from Figures \ref{fig:Mst_det} and \ref{fig:GF_mu} respectively. Refer to Table \ref{tab:fig4_table} for the average gas fractions from the scaling relations.}
\label{fig:gf-mu}
\end{figure*}

\begin{figure*}
\centering
  \captionsetup[subfigure]{labelformat=empty}
   \begin{subfigure}{0.49\linewidth} \centering
     \includegraphics[width=87mm,scale=1.5]{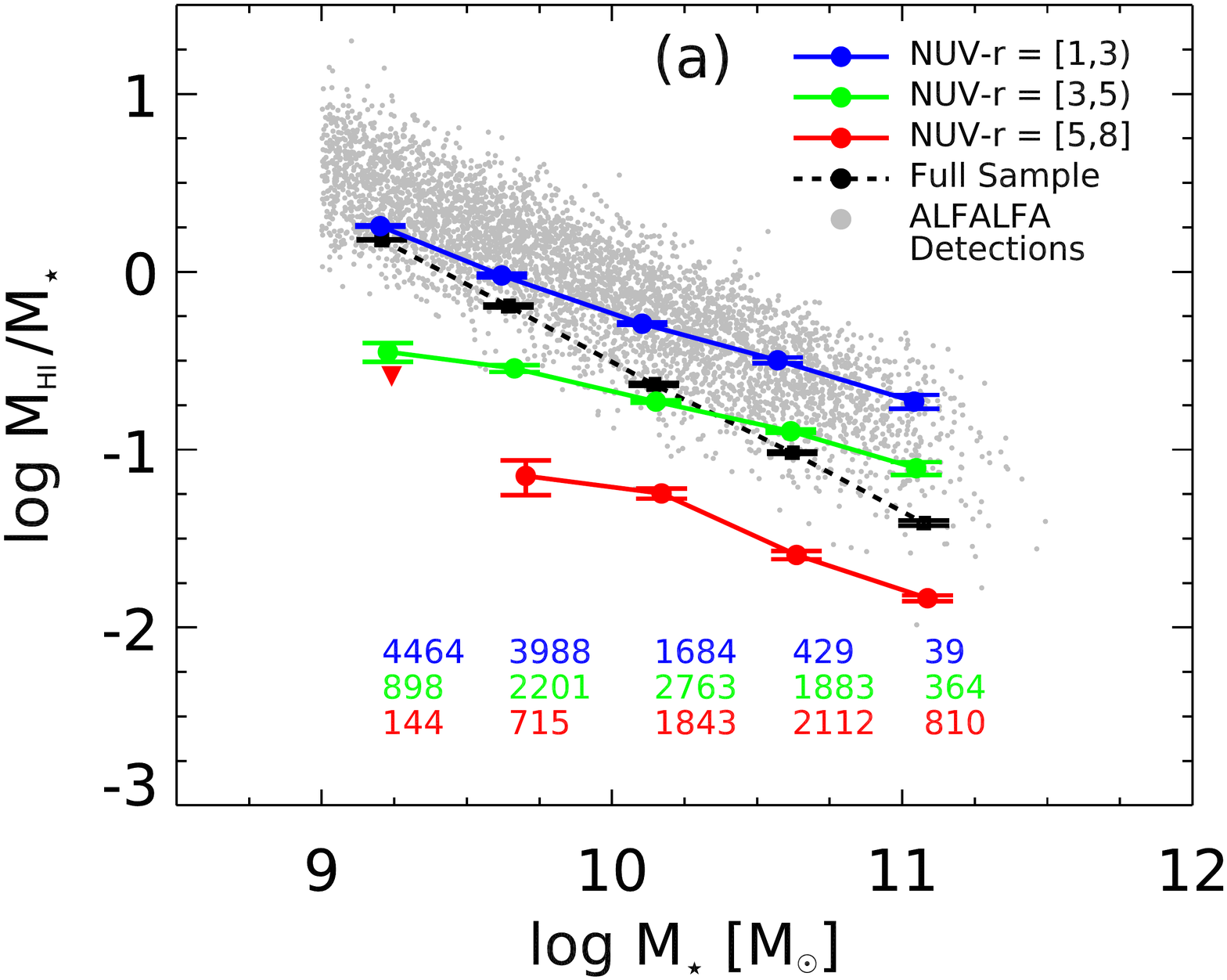}
     \caption{}\label{fig:figAnuvr}
   \end{subfigure}
   \begin{subfigure}{0.49\linewidth} \centering
     \includegraphics[width=87mm,scale=1.5]{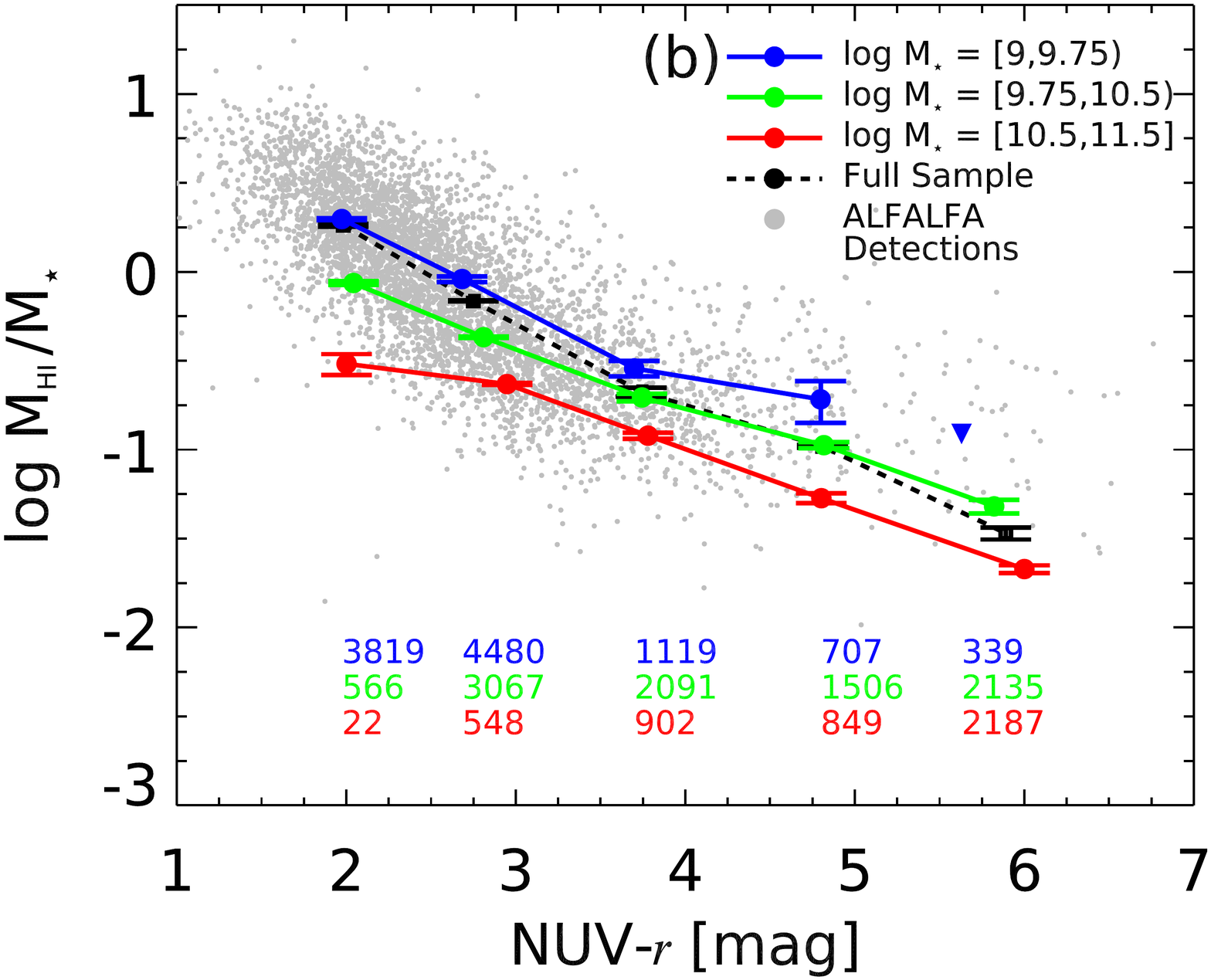}
     \caption{}\label{fig:figBnuvr}
   \end{subfigure}
\caption{Left: \HI gas fraction as a function of stellar mass, separated into bins of NUV-{\it r} colour, as indicated on the top right. Right: \HI gas fraction as a function of NUV-{\it r} colour; separated into bins of stellar mass. Symbols and numbers as in Figure \ref{fig:gf-mu}. Upper limits obtained for the bins where the stacked spectrum is a non-detection are shown as upside-down triangles. The dashed black lines reproduce gas fraction scaling relations for the whole sample from Figures \ref{fig:Mst_det} and \ref{fig:GF_nuvr} respectively. The gas fraction values are given in Table \ref{tab:fig5_table} for the average gas fractions for each scaling relation.}
\label{fig:gf-nuvr}
\end{figure*}

\begin{figure*}
\centering
  \captionsetup[subfigure]{labelformat=empty}
   \begin{subfigure}{0.49\linewidth} \centering
     \includegraphics[width=87mm,scale=1.5]{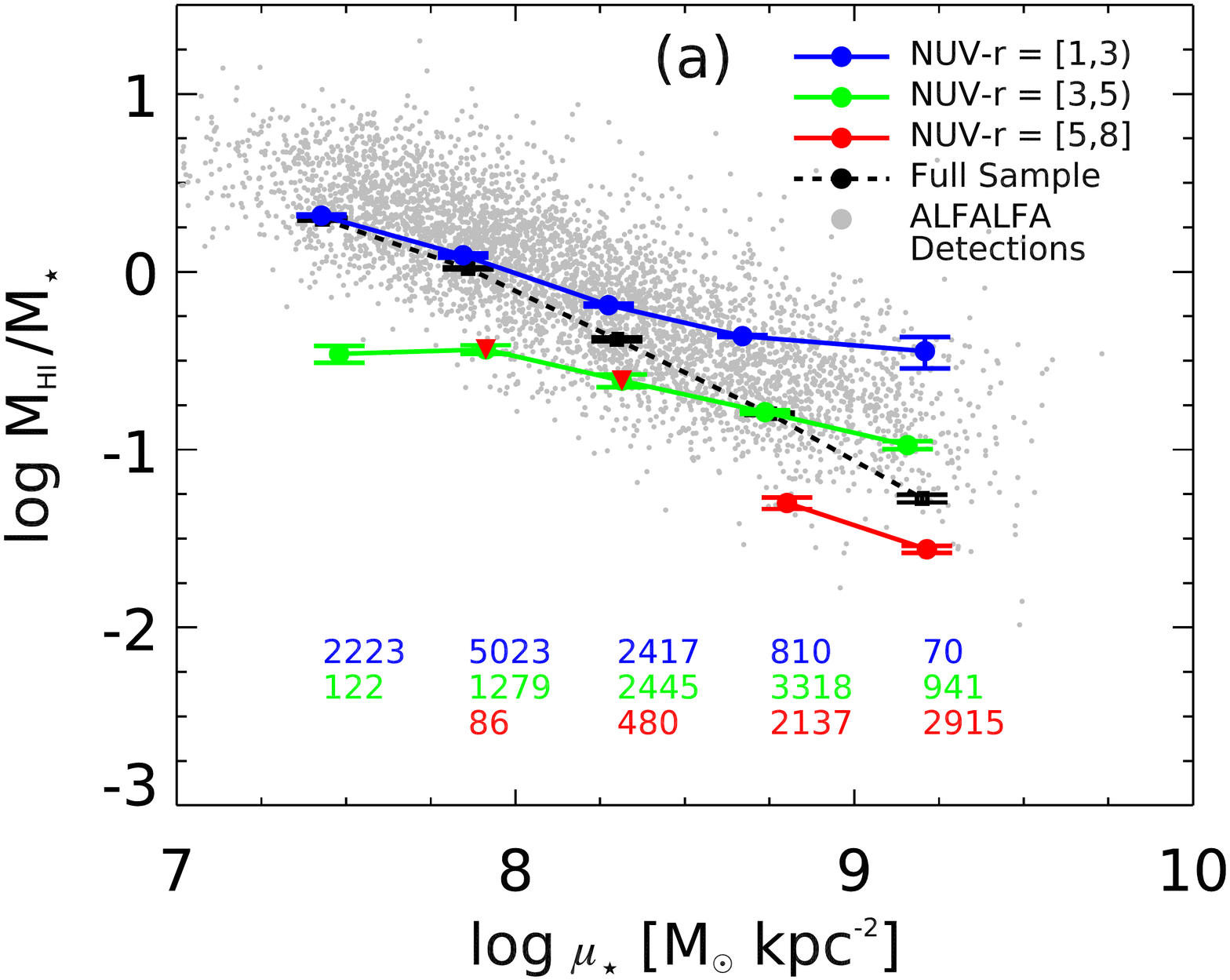}
     \caption{}\label{fig:figAmuNUVR}
   \end{subfigure}
   \begin{subfigure}{0.49\linewidth} \centering
     \includegraphics[width=87mm,scale=1.5]{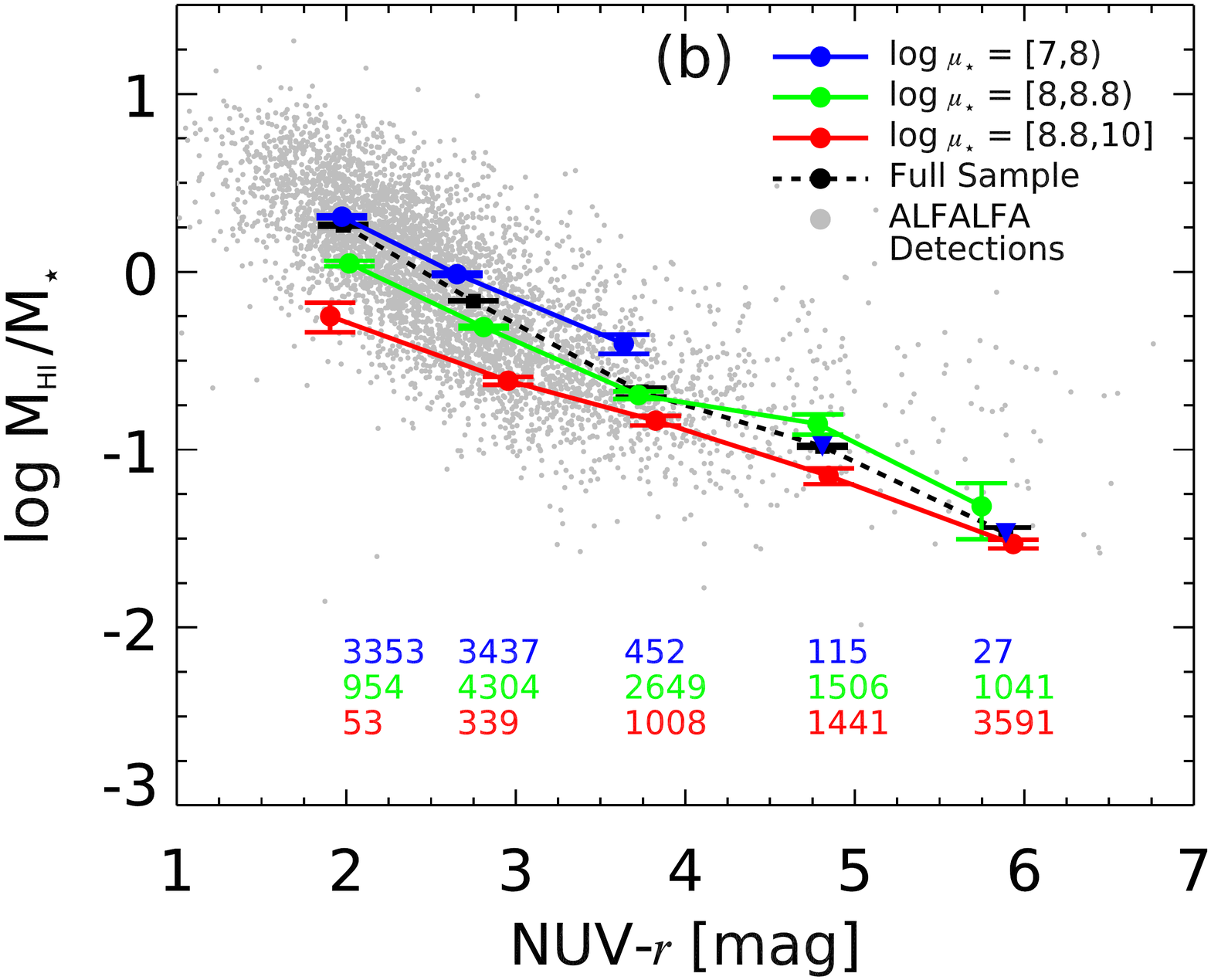}
     \caption{}\label{fig:figBNUVRmu}
   \end{subfigure}
\caption{Left:  Gas fraction as a function of  stellar surface density, separated into blue, green and red sequence galaxies according to their NUV-{\it r} colour (solid lines) as indicated on the top right. Right: Gas fraction as a function of NUV-{\it r} colour, separating into bins of stellar surface density. Symbols and numbers as in Figure \ref{fig:gf-mu}. The full sample scaling relations from Figures \ref{fig:GF_mu} and \ref{fig:GF_nuvr} are plotted as black dashed lines. Triangles denote non-detections, set to their upper limits. For the data points, refer to Table \ref{tab:fig6_table}.}
\label{fig:gf-mu-nuvr}
\end{figure*}

To take this analysis a step further, we must establish the importance of each of the main parameters as a tracer of gas content. Stellar surface density can be taken as a rough proxy for bulge-to-total ratio, a good morphological indicator, with the fraction of disk-dominated systems decreasing as \mustar \,increases \citep{Kauffmann2006}. Within our sample, more massive galaxies tend to have higher stellar surface densities and thus earlier morphologies. As the ratio of young to old stars, NUV-{\it r} colour is used as a proxy for specific star formation rate \citep[SSFR, see][]{Salim2005,Salim2007,Schiminovich2007}. We see in \citet{Cortese2011} and \citet{Catinella2013} that the \HI gas fraction is most tightly correlated with \mustar and NUV-{\it r} colour. However, no previous study has had the requisite amount of galaxies to divide the sample by two parameters simultaneously. The large statistics afforded by our sample allows us to fix a primary parameter while binning the sample in terms of a second, disentangling the individual dominance of  stellar mass, stellar surface density and NUV-{\it r} colour in governing, or at least tracing, the average gas fraction of these galaxies.

Figure \ref{fig:figAmu} shows gas fraction versus stellar mass, fixing stellar mass along the x-axis and splitting the sample into bins of stellar surface density. We then invert this in Figure \ref{fig:figBmu} so that stellar surface density is held constant and we are separating galaxies according to their stellar mass. Blue, green and red lines denote the average gas fractions obtained by stacking. For comparison, the scaling relations for the whole sample from Figure \ref{fig:main_SR} are shown by the dashed black line. Figure \ref{fig:figAmu} shows a large difference (\squiggle0.8 dex) between average \HI fraction for disk-dominated (low $\mu_{\star}$) and bulge-dominated (high $\mu_{\star}$) galaxies at a given stellar mass. In contrast, Figure \ref{fig:figBmu} has a smaller variation (\squiggle0.4 dex) in gas content across the sample's mass range while holding stellar surface density constant, suggesting that the average \HI fraction of a particular morphological class is, to a small degree, sensitive to the mass of the system. Using targeted \HI observations, \citet{Catinella2013} found that the distributions of gas fractions averaged in the stellar mass and stellar surface density relations have a standard deviation $\sigma_{M_{\star}} = $ 0.5 dex and $\sigma_{\mu_{\star}} = $ 0.4 dex respectively. As the spread in gas fraction between our second parameter bins is generally comparable to or larger than these values, we can safely conclude that the differences in gas fraction between the solid lines in Figure \ref{fig:gf-mu} are significant.

Lastly, we note that the average gas fraction in the lowest stellar mass ($\text{M}_{\star} < 10^{9.75} \text{M}_{\odot}$) and highest surface density bin in Figure \ref{fig:figBmu} is markedly above the relation. The higher than expected gas fraction is likely due to the uncertainty involved in calculating stellar surface densities for the large fraction of compact objects within the bin, over estimating the stellar surface densities. When statistics is low these galaxies dominate the average gas fraction measurement and the point should not be considered reliable.

Similarly, in Figure \ref{fig:gf-nuvr} we split the sample by NUV-{\it r} colour instead of stellar surface density. The blue, green and red lines denote galaxies within the NUV-{\it r} colour bins -  chosen to approximately correspond to traditional blue cloud, green valley and red sequence classifications - and stellar mass bins - chosen to span the transition mass of $10^{10 }\text{M}_{\odot}$ \citep{Kauffmann2003}, where there is an observed shift in the abundance of late, star forming galaxies to earlier, quiescent systems. Stacked spectra that remain undetected (upside-down triangles) are found exclusively on the red sequence with NUV-{\it r} $>\; 4.5$. Non-detections are set to their upper limits (see Section \ref{sec:stacking}).

Figure \ref{fig:figAnuvr} shows that for a given stellar mass the gas fraction varies significantly (\squiggle1.0 dex) across the NUV-{\it r} colour range, while in Figure \ref{fig:figBnuvr} the difference between average gas fraction at fixed NUV-{\it r} colour for low and high mass systems is significantly less (\squiggle0.5 dex). Even in this case, these differences are larger than the typical standard deviation of the scaling relation as directly measured from detections \citep[e.g. $\sigma_{NUV-r} = $ 0.3 dex,][]{Catinella2013}. The implication is that, once the trend of high stellar mass bins being dominated by redder galaxies is removed, the average \HI fraction for galaxies of given SSFR is only weakly dependent on mass.

When splitting the gas fraction-stellar mass scaling relation in terms of either surface density or NUV-{\it r} we find that the slope of the linear fit to the relation flattens considerably ($\nabla_{\mu_{\star}} = -0.45 \pm 0.01$; $\nabla_{\text{NUV-{\it r}}} = -0.35 \pm 0.02$) with respect to the whole sample ($\nabla_{\text{full}} = -0.85 \pm 0.01$) shown by the dashed black line in Figures \ref{fig:gf-mu} and \ref{fig:gf-nuvr}. This clearly shows that {\it the steep slope of the gas fraction-stellar mass relation arises as a result of preferentially stacking blue, gas-rich galaxy populations in low stellar mass bins and red, gas-poor systems in high mass bins.}

Having confirmed that stellar surface density and NUV-{\it r} colour dominate over stellar mass as tracers of atomic gas content, the next step is to test which one of these is the principal parameter in the determination of average \HI fraction. The best method to address this is presented in Figure \ref{fig:gf-mu-nuvr}, the main result of this work. We remove the stellar mass constraints and show how the gas content varies when we fix surface density or NUV-{\it r} colour while binning in terms of the other. As with Figures \ref{fig:gf-mu} and \ref{fig:gf-nuvr}, the non-detections occur exclusively in bins of NUV-{\it r} $>\;4.5$, where the stacking of large numbers of galaxies is required to reduce the {\it rms} noise sufficiently for the signal to be detected. Figure \ref{fig:figAmuNUVR} demonstrates that galaxies at a given stellar surface density exhibit a spread (\squiggle1.0 dex) in \HI content across the NUV-{\it r} colour range (solid lines) from blue cloud to red sequence. The difference of 1.0 dex is statistically significant when compared to the scatter in the gas fraction-stellar surface density relation found by \citet{Catinella2013}, $\sigma_{\mu_{\star}} = $ 0.4 dex. In contrast, we fix the colour in Figure \ref{fig:figBNUVRmu} while separating galaxies according to low, intermediate or high stellar surface density (solid lines). In this case the difference in gas fraction between the highest and lowest $\mu_{\star}$ bins, i.e. the bulge- and disk-dominated systems, decreases to \squiggle0.5 dex on average.

This implies that the most important quantity in tracing of neutral atomic hydrogen content is NUV-{\it r} colour. Making additional cuts in surface density, already a mass dependent quantity, does not significantly alter the values of gas fraction. In other words, galaxies of a similar colour are, on average, likely to contain similar ratios of cold gas to stellar mass, showing only a small dependence on size (e.g. mass) or morphology (e.g. surface density).

While the strong relation between gas and NUV-{\it r} colour is evident, Figures \ref{fig:figBnuvr} and \ref{fig:figBNUVRmu} do demonstrate a smaller, residual dependence of gas fraction on stellar mass and density at fixed NUV-{\it r} colour. This result raises interesting questions surrounding the impact of stellar mass and surface density on gas consumption in addition to star formation, which will be discussed in Section \ref{sec:Discussion}.

In addition, the slope of the gas fraction-surface density relation is flattened when binning galaxies by NUV-{\it r} colour. The linear fit to the full sample relation has a gradient of $\nabla_{\text{full}} = -0.89 \pm 0.04$, while selection by NUV-{\it r} reduces the slope to $\nabla_{\text{NUV-{\it r}}} = -0.33 \pm 0.06$. This reaffirms that selection by only surface density yields a mixed  population of galaxies and shows that the surface density scaling relation is driven by the underlying correlation of gas content with NUV-{\it r} colour.

To confirm the low affinity of gas fraction with stellar mass, we also split the scaling relations of Figure \ref{fig:gf-mu-nuvr} into lower ($\text{M}_{\star} < 10^{10}\; \text{M}_{\odot}$) and higher stellar mass ($\text{M}_{\star} \geq 10^{10}\; \text{M}_{\odot}$) bins (not shown). The difference between the relations of Figure \ref{fig:gf-mu-nuvr} for low and high mass galaxies was not significant, \squiggle0.1 dex (a) and \squiggle0.2 dex (b). This reaffirms that, within our sample, the variation of gas content across the stellar mass range is small, once morphology and/or SSFR of galaxies has been fixed.

Finally, we stress that by looking at the effect of secondary parameters on the \HI scaling relations we have been able to put important constraints on the underlying gas fraction distribution of the stacked population. In other words, we have been able to overcome one of the main limitations of \HI stacking, thus enhancing the scientific potential of this technique. 

\section{Discussion and Conclusions}
\label{sec:Discussion}
In this work we have applied \HI spectral stacking to a large sample of $24,337$ galaxies. Each galaxy is selected according to redshift and stellar mass from the SDSS DR7, with \HI spectra and NUV data from the ALFALFA blind \HInospace-survey and GALEX catalogues respectively. The goal of this study is to investigate the dependence of gas content on the galaxy properties of stellar mass, stellar surface density and NUV-{\it r} colour, extending the previous work of \citet{Fabello2011a} down to lower stellar masses ($\text{M}_{\star} \geq 10^{9} \text{M}_{\odot}$) and significantly increasing the number of galaxies.

The key conclusions of this paper can be can be summarised as follows:

\begin{enumerate}
  \item[1.] We confirm that NUV-r colour excels over stellar mass and stellar surface density as a tracer of galaxy gas content, as previously noted by \citet{Cortese2011}, \citet{Fabello2011a} and \citet{Catinella2013}. Additionally, we quantify the strong decrease of gas fraction with increasing NUV-{\it r} colour at fixed stellar mass and stellar surface density.
  
  \item[2.] We show for the first time that the gas fraction-stellar mass and, to first order, the gas fraction-surface density scaling relations are driven by the primary correlation of gas content with NUV-{\it r} colour.

  \item[3.] At fixed NUV-{\it r} colour we find a small residual dependence on stellar mass and stellar surface density. This suggests a residual effect of mass and morphology on gas consumption at fixed SSFR, as discussed below.
\end{enumerate}

As already mentioned, one may regard NUV-{\it r} colour as a proxy for SSFR and thus, under the simple assumption that gas fraction and colour are {\it derived over the same surface area}, it is easy to show that the gas fraction-NUV-{\it r} relation can be interpreted as an integrated \HI Kennicutt-Schmidt (KS) law \citep{Schmidt1959,Kennicutt1998} relating the atomic gas content of galaxies to their star formation activity. In this context, our findings not only confirm that star formation is the property of galaxies most closely related to their \HI gas content, but also highlight that the main scaling relations of gas fraction-stellar mass, -stellar surface density and -NUV-{\it r} colour can be understood as a combination of the underlying bimodality in specific star formation and the KS relation. Low mass or disk galaxies are preferentially blue, star formers, whereas massive or bulge-dominated systems are, on average, more quiescent. This means that the gas fraction-stellar mass and -stellar surface density relations are simply driven by the variation in SSFR between the two populations.

Given that NUV-{\it r} colour is the principle driver behind the main integrated gas fraction scaling relations, it is interesting that we find a residual dependence on stellar mass and surface density in the gas-fraction-NUV-{\it r} plane (see Figures \ref{fig:figBnuvr} and \ref{fig:figBNUVRmu}). High mass and surface density galaxies have lower \HI fraction than low surface density objects at fixed NUV-{\it r} colour. If, as above, we take NUV-{\it r} as equivalent to SSFR, this shows that massive and bulge-dominated galaxies have a lower gas fraction than low mass and disk-dominated systems respectively for the same SSFR. It follows from this that the timescales over which gas is consumed by star formation (hereafter depletion time, M$_{\text{HI}}/$SFR) are shorter and therefore star formation efficiency (SFE $\equiv$ SFR$/$M$_{\text{HI}}$) is higher in more massive or bulge-dominated galaxies than their low mass or disk-dominated counterparts, assuming that star formation continues at its current rate. Our evidence contributes to a physical picture where the gas content of galaxies is strongly regulated by star formation and, albeit to a lesser extent, influenced by both mass and structure. This is tentative evidence that the KS relation is to some degree dependent upon the morphological properties of the galaxy. 

Assuming that our results might extend to the molecular gas (H$_{2}$) component, the secondary dependence of gas content on stellar mass and surface density is in qualitative agreement with \citet{Huang2014b} who examined the variation in H$_{2}$ depletion times in bulges, spirals, bars and rings, at fixed SSFR, for a sample of massive ($\text{M}_{\star} > 10^{10}\; \text{M}_{\odot}$) local galaxies. Their results show that, at fixed SSFR, the \Htwo depletion times are shorter for bulge-dominated galaxies. In their discussion they invoke the conclusions of \citet{Helfer1993} as a possible explanation, whereby gravitational potential and the density of molecular clouds is increased in the presence of a stellar bulge.

\begin{figure}
\centering
\hspace{-13mm}
\includegraphics[width=90mm,scale=1.5]{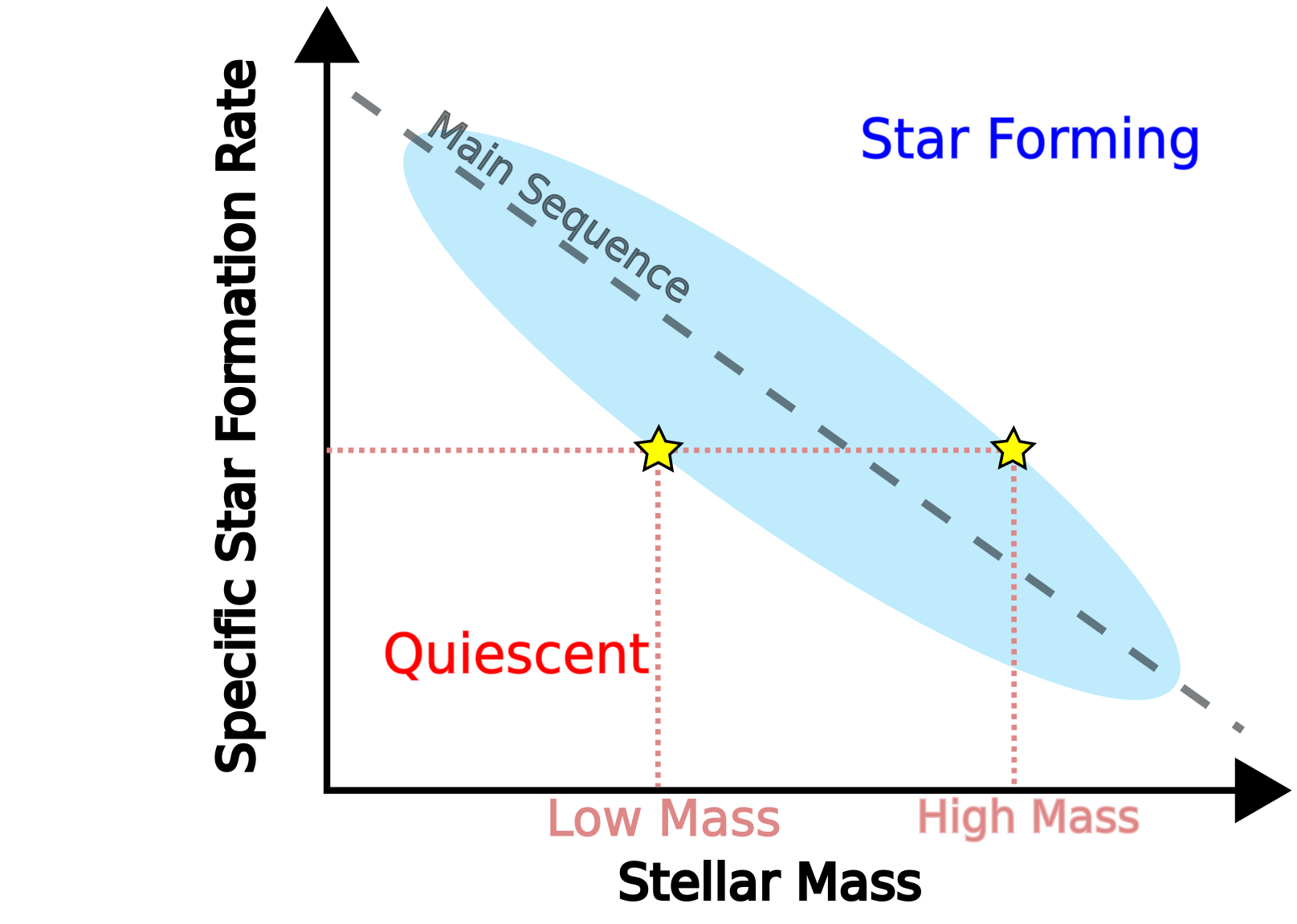} % Include the image
\caption{Diagram outlining the dependence of SSFR on stellar mass. The scatter in the SSFR-\Mstar relation is represented by the blue ellipse and the grey-blue dashed line denotes the star forming `main sequence'. Galaxies classified as quiescent reside below the main sequence while star forming galaxies are found above. The pink dotted line shows a constant value of SSFR; the galaxies lying `below' the main sequence, and thus deemed quiescent, are low mass systems while the high mass galaxies deviate `above' the main sequence in their role as star formers.}
\label{fig:SF-MS}
\end{figure}

Conversely, \citet{Saintonge2012} find that for their sample of galaxies, a subset of which is analysed in \citet{Huang2014b}, the shortest depletion times and thus the highest SFEs are found in disk-dominated galaxies. This conclusion is reached by examining SFE as a function of distance from the star formation (SF) main sequence, showing that weakly star forming galaxies have low gas content and long depletion times, and combining this with the derived \Htwo KS law where high surface density systems lie systematically below the mean relation.

At face value, our result showing variation in gas content with surface density seems contradictory to the conclusion of \citet{Saintonge2012}. However, it is easy to show that the inconsistency is only apparent. Both \citet{Huang2014b} and this work examine depletion time at fixed SSFR while, on the other hand, \citet{Saintonge2012} leave SSFR unconstrained and investigate depletion time as a function of distance from the SF main sequence. Once one accounts for the difference in method our findings are entirely consistent with those of \citet{Saintonge2012}. To illustrate this, the cartoon in Figure \ref{fig:SF-MS} shows that for fixed SSFR, high mass galaxies are clearly offset above the specific SF main sequence \citep[see][]{Salim2007} while low mass galaxies are found in the quiescent region below the mean relation. The increasing prominence of bulges in high mass galaxies leads to bulge-dominated galaxies being deemed {\it stronger star formers} in comparison with disks at fixed SSFR. Accepting this, Figure \ref{fig:SF-MS} illustrates how bulge-dominated galaxies lie above the specific SF main sequence and thus must have lower depletion time in comparison to disks at fixed SSFR, a result that is in agreement with the works of both \citet{Huang2014b} and \citet{Saintonge2012}.

Of course it is important to offer more conclusive arguments on the relationship between gas and star formation in galaxies and to do so we must determine the physical star formation rates and efficiencies, rather than employing the proxy of NUV-{\it r} colour. This will be part follow-up papers.

Our work introduces a more complete description of the relationships between galaxy properties and gas content, an area that provides strong constraints for galaxy formation and evolution models by disentangling the influence of these properties on gas content. We demonstrate the importance and power of \HI spectral stacking, a technique with great potential for investigating the physical mechanisms that drive the evolution of galaxies by probing the cold gas content of huge galaxy samples in regimes that would otherwise be inaccessible until the full capability of the Square Kilometre Array is realised.

\section*{Acknowledgments}

We wish to acknowledge the work of the entire ALFALFA team in observing, flagging, and processing the ALFALFA data that this paper makes use of. 

BC is the recipient of an Australian Research Council Future Fellowship (FT120100660). This research was supported under the Australian Research Council's Discovery Projects funding scheme (DP130100664 and DP150101734).

RG and MPH are supported by NSF grant AST 1107390 and by the Brinson Foundation.

GALEX (Galaxy Evolution Explorer) is a NASA Small Explorer, launched in April 2003. We gratefully acknowledge NASA's support for construction, operation, and science analysis for the GALEX mission, developed in cooperation with the Centre National d'Etudes Spatiales (CNES) of France and the Korean Ministry of Science and Technology.

Funding for SDSS-III has been provided by the Alfred P. Sloan Foundation, the Participating Institutions, the National Science Foundation, and the U.S. Department of Energy Office of Science. The SDSS-III web site is \url{http://www.sdss3.org/}.

SDSS-III is managed by the Astrophysical Research Consortium for the Participating Institutions of the SDSS-III Collaboration including the University of Arizona, the Brazilian Participation Group, Brookhaven National Laboratory, Carnegie Mellon University, University of Florida, the French Participation Group, the German Participation Group, Harvard University, the Instituto de Astrofisica de Canarias, the Michigan State/Notre Dame/JINA Participation Group, Johns Hopkins University, Lawrence Berkeley National Laboratory, Max Planck Institute for Astrophysics, Max Planck Institute for Extraterrestrial Physics, New Mexico State University, New York University, Ohio State University, Pennsylvania State University, University of Portsmouth, Princeton University, the Spanish Participation Group, University of Tokyo, University of Utah, Vanderbilt University, University of Virginia, University of Washington, and Yale University.

% this section writes out the bibliography section
% mnras is the style: requires mnras.bst
\bibliographystyle{mn2e}
% refs.bib is the name of your bibtex database
\bibliography{refs}

% \begin{landscape}
% \appendix

% \section[]{Appendix A: Average Gas Fractions}
%%%%%%%%%%%%%%%%%%%%%%%%%%%%%%%%%%%%%%%%%%%%%%%%%%%%%%%%%%
\begin{table}
\caption{Average gas fractions for the scaling relations shown in Figure \ref{fig:gf-mu}. The column labelled {\it y} is the secondary property and limits by which we bin the sample.}
\label{tab:fig4_table}
\begin{tabular}{lcrll}
\hline

{\it x}    & {\it y}    & $\langle x \rangle$    & $\langle \text{M}_{\text{HI}}/\text{M}_{\star} \rangle$    &   N\\

\hline

log M$_{\star}$    &    $7 \; \leq$ log $\mu_{\star}$ $< \; 8$    &    9.20  &  1.722 $\pm$ 0.029  &  4204 \\
                             &                                                                  &   9.59  &  1.000 $\pm$ 0.017  &  2808 \\
                             &                                                                  &   10.02  &  0.589 $\pm$ 0.018  &  363 \\

                             &                                                                  &              &                                  &          \\
                             &                                                                  &              &                                  &          \\
                             &                                                                  &              &                                  &          \\                                                          

                             &    $8 \; \leq$ log $\mu_{\star}$ $< \; 8.8$  &   9.24  &  0.735 $\pm$ 0.030  &  1226 \\
                             &                                                                   &   9.67  &  0.429 $\pm$ 0.008  &  3733 \\
                             &                                                                   &   10.13  &  0.275 $\pm$ 0.004  &  3888 \\
                             &                                                                   &   10.58  &  0.174 $\pm$ 0.002  &  1488 \\
                             &                                                                   &   11.03  &  0.114 $\pm$ 0.008  &  119 \\

                             &                                                                   &              &                                  &          \\

                             &    $8.8 \; \leq$ log $\mu_{\star}$ $\leq \; 10$    &   9.75  &  0.168 $\pm$ 0.029  &  360 \\
                             &                                                                           &   10.19  &  0.095 $\pm$ 0.008  &  2038 \\
                             &                                                                           &   10.64  &  0.056 $\pm$ 0.001  &  2924 \\
                             &                                                                           &   11.08  &  0.032 $\pm$ 0.002  &  1092 \\

                             &                                                                           &              &                                  &          \\
                             &                                                                           &              &                                  &          \\
                                &                                 &              &                                  &          \\

log $\mu_{\star}$    &    $9 \; \leq$ log M$_{\star}$ $< \; 9.75$    &   7.43  &  1.990 $\pm$ 0.028  &  2328 \\
                                &                                                                    &   7.84  &  1.151 $\pm$ 0.010  &  5070 \\
                                &                                                                    &   8.27  &  0.538 $\pm$ 0.016  &  2222 \\
                                &                                                                    &   8.67  &  0.288 $\pm$ 0.016  &  740 \\
                                &                                                                    &   9.13  &  0.476 $\pm$ 0.120  &  44 \\

                                &                                 &              &                                  &          \\

                                &    $9.75 \; \leq$ log M$_{\star}$ $< \; 10.5$    &   7.53  &  0.993 $\pm$ 0.088  &  21 \\
                                &                                                                         &   7.95  &  0.622 $\pm$ 0.020  &  1314 \\
                                &                                                                         &   8.31  &  0.328 $\pm$ 0.007  &  2868 \\
                                &                                                                         &   8.74  &  0.159 $\pm$ 0.009  &  3638 \\
                                &                                                                         &   9.18  &  0.068 $\pm$ 0.007  &  1522 \\

                                &                                 &              &                                  &          \\

                                &    $10.5 \; \leq$ log M$_{\star}$ $\leq \; 11.5$    &   8.38  &  0.238 $\pm$ 0.012  &  252 \\
                                &                                                                             &   8.80  &  0.098 $\pm$ 0.002  &  1887 \\
                                &                                                                             &   9.21  &  0.037 $\pm$ 0.002  &  2360 \\

                                &                                 &              &                                  &          \\
                                &                                 &              &                                  &          \\
                                &                                 &              &                                  &          \\

\hline
\end{tabular}
\end{table}

%%%%%%%%%%%%%%%%%%%%%%%%%%%%%%%%%%%%%%%%%%%%%%%%%%%%%%%%%%

\begin{table}
\caption{Average gas fractions for the scaling relations shown in Figure \ref{fig:gf-nuvr}. Numbers preceded by a ``$<$'' sign are upper limits.\\}
\label{tab:fig5_table}
\begin{tabular}{lcrll}
\hline

{\it x}    & {\it y}    & $\langle x \rangle$    &  $\langle \text{M}_{\text{HI}}/\text{M}_{\star} \rangle$     &   N\\

\hline

log M$_{\star}$    &    $1 \; \leq$ NUV-{\it r} $< \; 3$    &    9.20  &  1.805 $\pm$ 0.017  &  4464 \\
                             &                                                         &   9.62  &  0.953 $\pm$ 0.021  &  3988 \\
                             &                                                         &   10.10  &  0.510 $\pm$ 0.008  &  1684 \\
                             &                                                         &   10.57  &  0.318 $\pm$ 0.012  &  429 \\
                             &                                                         &   11.04  &  0.186 $\pm$ 0.017  &  39 \\

                             &                                                                   &              &                                  &          \\

                             &    $3 \; \leq$ NUV-{\it r} $< \; 5$    &   9.23  &  0.355 $\pm$ 0.040  &  898 \\
                             &                                                         &   9.66  &  0.286 $\pm$ 0.013  &  2201 \\
                             &                                                         &   10.15  &  0.186 $\pm$ 0.003  &  2763 \\
                             &                                                         &   10.62  &  0.127 $\pm$ 0.003  &  1883 \\
                             &                                                         &   11.05  &  0.079 $\pm$ 0.007  &  364 \\

                             &                                                                   &              &                                  &          \\

                             &    $5 \; \leq$ NUV-{\it r} $\leq \; 8$    &   9.24  &  \multicolumn{1}{c}{$<$ 0.258}  &  144 \\
                             &                                                         &   9.70  &  0.071 $\pm$ 0.016  &  715 \\
                             &                                                         &   10.17  &  0.057 $\pm$ 0.004  &  1843 \\
                             &                                                         &   10.64  &  0.026 $\pm$ 0.001  &  2112 \\
                             &                                                         &   11.09  &  0.015 $\pm$ 0.001  &  810 \\

                             &                                 &              &                                  &          \\
                             &                                 &              &                                  &          \\

NUV-{\it r}           &    $9 \; \leq$ log M$_{\star}$ $< \; 9.75$    &    1.97  &  1.979 $\pm$ 0.028  &  3819 \\
                                                        &                                         &     2.68  &  0.910 $\pm$ 0.033  &  4480 \\
                                                        &                                         &     3.70  &  0.286 $\pm$ 0.029  &  1119 \\
                                                        &                                         &     4.80  &  0.192 $\pm$ 0.051  &  707 \\
                                                        &                                         &     5.63  &  \multicolumn{1}{c}{$<$ 0.123} &  339 \\

                                &                                 &              &                                  &          \\

                        &    $9.75 \; \leq$ log M$_{\star}$ $< \; 10.5$    &     2.04  &  0.865 $\pm$ 0.023  &  566 \\
                                                        &                                         &     2.81  &  0.429 $\pm$ 0.005  &  3067 \\
                                                        &                                         &     3.74  &  0.196 $\pm$ 0.010  &  2091 \\
                                                        &                                         &     4.82  &  0.106 $\pm$ 0.004  &  1506 \\
                                                        &                                         &     5.82  &  0.048 $\pm$ 0.004  &  2135 \\

                                &                                 &              &                                  &          \\

                    &    $10.5 \; \leq$ log M$_{\star}$ $\leq \; 11.5$    &     2.00  &  0.304 $\pm$ 0.041  &  22 \\
                                                        &                                         &     2.95  &  0.234 $\pm$ 0.003  &  548 \\
                                                        &                                         &     3.78  &  0.120 $\pm$ 0.005  &  902 \\
                                                        &                                         &     4.80  &  0.053 $\pm$ 0.003  &  849 \\
                                                        &                                         &     6.00  &  0.021 $\pm$ 0.001  &  2187 \\

                                &                                 &              &                                  &          \\

\hline                                
\end{tabular}
\end{table}

%%%%%%%%%%%%%%%%%%%%%%%%%%%%%%%%%%%%%%%%%%%%%%%%%%%%%%%%%%
\newpage
\newpage

\begin{table}
\caption{Average gas fractions for the scaling relations shown in Figure \ref{fig:gf-mu-nuvr}. Upper limits are preceded by a ``$<$'' sign.}
\label{tab:fig6_table}

\begin{tabular}{lcrll}
\hline

{\it x}    & {\it y}    & $\langle x \rangle$    & {\bf $\langle \text{M}_{\text{HI}}/\text{M}_{\star} \rangle$ }   &   N\\

\hline

log $\mu_{\star}$    &    $1 \; \leq$ NUV-{\it r} $< \; 3$    &   7.43  &  2.069 $\pm$ 0.026  &  2223 \\
                                &                                                        &   7.85  &  1.238 $\pm$ 0.035  &  5023 \\
                                &                                                        &   8.27  &  0.649 $\pm$ 0.012  &  2417 \\
                                &                                                        &   8.67  &  0.435 $\pm$ 0.006  &  810 \\
                                &                                                        &   9.21  &  0.358 $\pm$ 0.072  &  70 \\

                                &                                 &              &                                  &          \\

                                &    $3 \; \leq$ NUV-{\it r} $< \; 5$    &   7.48  &  0.346 $\pm$ 0.038  &  122 \\
                                &                                                        &   7.91  &  0.366 $\pm$ 0.021  &  1279 \\
                                &                                                        &   8.31  &  0.244 $\pm$ 0.020  &  2445 \\
                                &                                                        &   8.74  &  0.162 $\pm$ 0.004  &  3318 \\
                                &                                                        &   9.16  &  0.106 $\pm$ 0.005  &  941 \\

                                &                                 &              &                                  &          \\

                                &    $5 \; \leq$ NUV-{\it r} $\leq \; 8$    &   7.95  &  \multicolumn{1}{c}{$<$ 0.318} &  86 \\
                                &                                                        &   8.35  &  \multicolumn{1}{c}{$<$ 0.089}  &  480 \\
                                &                                                        &   8.80  &  0.050 $\pm$ 0.004  &  2137 \\
                                &                                                        &   9.21  &  0.028 $\pm$ 0.001  &  2915 \\

                                &                                 &              &                                  &          \\
                                &                                 &              &                                  &          \\
                                                                &                                 &              &                                  &          \\

NUV-{\it r}           &    $7 \; \leq$ log $\mu_{\star}$ $< \; 8$    &    1.97  &  2.040 $\pm$ 0.037  &  3353 \\
                             &                                                                  &   2.65  &  0.969 $\pm$ 0.021  &  3437 \\
                             &                                                                  &   3.64  &  0.394 $\pm$ 0.049  &  452 \\
                             &                                                                  &   4.71  &  \multicolumn{1}{c}{$<$ 0.294}  &  115 \\
                             &                                                                  &   5.69  &  \multicolumn{1}{c}{$<$ 0.512}  &  27 \\

                              &                                 &              &                                  &          \\

                             &    $8 \; \leq$ log $\mu_{\star}$ $< \; 8.8$  &   2.02  &  1.114 $\pm$ 0.041  &  954 \\
                             &                                                                  &   2.81  &  0.489 $\pm$ 0.009  &  4304 \\
                             &                                                                  &   3.72  &  0.203 $\pm$ 0.009  &  2649 \\
                             &                                                                  &   4.78  &  0.140 $\pm$ 0.018  &  1506 \\
                             &                                                                  &   5.75  &  0.048 $\pm$ 0.017  &  1041 \\

                                &                                 &              &                                  &          \\

                             &   $8.8 \; \leq$ log $\mu_{\star}$ $\leq \; 10$  &   1.91  &  0.562 $\pm$ 0.110  &  53 \\
                             &                                                                  &   2.95  &  0.244 $\pm$ 0.013  &  339 \\
                             &                                                                  &   3.83  &  0.146 $\pm$ 0.009  &  1008 \\
                             &                                                                  &   4.85  &  0.071 $\pm$ 0.007  &  1441 \\
                             &                                                                  &   5.93  &  0.030 $\pm$ 0.002  &  3591 \\

                                &                                 &              &                                  &          \\
                                &                                 &              &                                  &          \\

\hline
\end{tabular}
\end{table}

\label{lastpage}

\end{document}